\documentclass[12pt]{article}
\usepackage{latexsym,graphics,graphicx,amssymb,amsmath, color}
\makeatletter
\title{}
\author{}
\date{}
\newcommand{\be}{\begin{equation}}
\newcommand{\ee}{\end{equation}}

\newcommand{\bc}{\begin{center}}
\newcommand{\ec}{\end{center}}
\newcommand{\bfl}{\begin{flushleft}}
\newcommand{\efl}{\end{flushleft}}

\newcommand{\beqa}{\begin{eqnarray}}
\newcommand{\eeqa}{\end{eqnarray}}
\newcommand{\beqan}{\begin{eqnarray*}}
\newcommand{\eeqan}{\end{eqnarray*}}
\newcommand{\beq}{\begin{equation}}
\newcommand{\eeq}{\end{equation}}
\newcommand{\prf}{\noindent {\bf Proof}\ \ \ }

\newcommand{\lbr}{\left \{ }
\newcommand{\rbr}{\right \} }
\newcommand{\Lbr}{\left [}
\newcommand{\Rbr}{\right ]}
\newcommand{\lp}{\left (}
\newcommand{\rp}{\right )}
\newcommand{\df}{\stackrel{{\rm def}}{=}}
\newcommand{\real}{{\mathbb{R}}}

\newcommand{\ints}{{\mathbb{Z}}}
\newcommand{\prob}{{\mathbb{P}}}

\newcommand{\sgn}{{\mathrm{sgn}}}
\newcommand{\E}{{\mathbb{E}}}
\newcommand{\C}{\mathcal}
\newcommand{\ve}[1]{{\bf {#1}}}
\newcommand{\m}[1]{{\bf {#1}}}
\newcommand{\Co}{\mathrm{co}}
\newcommand{\Var}{\mathrm{Var}}
\newtheorem{prop}{Proposition}
\newtheorem{lemma}{Lemma}
\newtheorem{theorem}{Theorem}
\newtheorem{claim}{Claim}
\newcommand{\markov}{\leftrightarrow}

\author{Saurabha Tavildar, Pramod Viswanath, and Aaron B.\ Wagner}
\title{The Gaussian Many-Help-One Distributed Source Coding
Problem}
\date{19 January, 2008}

\begin{document}
\maketitle

\begin{abstract}
Jointly Gaussian memoryless sources are observed at  $N$ distinct
terminals. The goal is to efficiently encode the observations in a
distributed fashion so as to enable reconstruction of any one of the
observations, say the first one, at the decoder subject to a
quadratic fidelity criterion. Our main result is a {\em precise}
characterization of the rate-distortion region when the covariance
matrix of the sources satisfies a ``tree-structure'' condition. In
this situation,  a natural analog-digital separation scheme
optimally trades off the distributed quantization rate tuples and
the distortion in the reconstruction: each encoder consists of a
point-to-point Gaussian vector quantizer followed by a Slepian-Wolf binning
encoder. We also provide a partial converse that suggests that the
tree structure condition is fundamental.
\end{abstract}

\section{Introduction}

\begin{figure}[ht]
\begin{center}
\scalebox{1}{\input{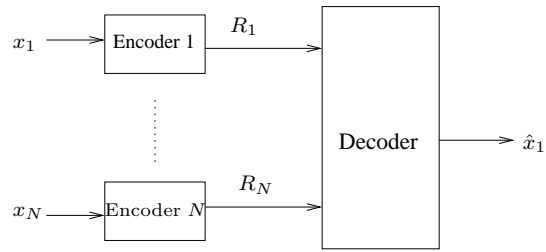}}
\end{center}
\caption{The many-help-one problem.} \label{fig:manyhelpone}
\end{figure}

The focus of this  study is the problem of distributed source coding of
 memoryless Gaussian sources with quadratic distortion constraints. The
rate-distortion region of this problem with two terminals has been recently
characterized \cite{WTV06}. Our focus, hence, is on the case when there are
at least 3 terminals. In this paper, we study a special case of this general
problem: the so-called ``many-help-one'' situation depicted in
Figure~\ref{fig:manyhelpone}.
The setup is the following:
\begin{itemize}
\item {\em Sources}:
Each of the $N$ encoders observes a memoryless discrete-time source:
encoder $i$ observes, over $n$ discrete time instants, the
memoryless source $x_i^n$. The observations across the encoders are
correlated, however. Specifically, the joint observations at time
$m$ $\lp x_1(m),\ldots, x_N(m)\rp$ are jointly Gaussian. Further,
the joint observations are memoryless over time $m$.
\item {\em Encoders}:
Each  encoder $i$ maps the vector of analog observations (over $n$ time instants, say) into a
vector of bits  (of length $R_i n$, say) that is then communicated without loss
to a single decoder (on a link with rate $R_i$).
\item {\em Decoder}:
The decoder is only interested in reconstructing one of the sources, 
(say,  $x_1^n$).
The fidelity criterion considered here is  a quadratic one: the average (over
the statistics of the sources) $l_2$ distance
between the original source vector and the reconstructed vector is required to
be no more than $Dn$.
\item {\em Problem statement}: The problem is to characterize the minimum set of rates at
which the encoders can communicate with the decoder while still conveying enough information
to satisfy the quadratic distortion constraint on the reconstruction.
\end{itemize}

In this paper, we precisely characterize the rate-distortion region
of a class of many-help-one problems. A crucial step towards solving
this problem involves the introduction of  a related distributed
source coding problem where the source has a ``binary tree'' structure;
this is done in  Section \ref{sec:binary}.
 We show that the natural analog-digital separation
strategy of point-to-point Gaussian vector quantization followed by a
distributed Slepian-Wolf binning scheme is optimal for this problem
(this is done in Sections~\ref{sec:tree_ach} and
\ref{sec:tree_conv}). Next, we show how this  result can be used to
solve various instances of the  many-help-one problem of interest; this
is done in Section \ref{sec:manyhelpone}. Finally, various ancillary aspects of the
problem at hand are discussed in Section~\ref{sec:discussion}: specifically the worst-case property of the Gaussian distribution with respect to 
the analog-digital separation architecture is demonstrated and a partial
 converse for the necessity of the tree-structure condition is provided.

\section{The Binary Tree Structure Problem}\label{sec:binary}
In this section, we take a short detour away from the many-help-one
problem of interest (c.f.\ Figure~\ref{fig:manyhelpone}).
Specifically, we introduce a related distributed source coding
problem that we call the ``binary tree structure problem".  We show
that the natural analog-digital separation architecture is
optimal in terms of the rate-distortion tradeoff
for this problem. The connection between the original
many-help-one problem and this binary tree structure problem is
made in the next section.

The outline of this section is as follows:
\begin{itemize}
\item we introduce the source variables and their statistical
relationships first (Section~\ref{sec:tree});
\item next we specify precisely the binary tree structure problem
 (Section~\ref{sec:binarytreeproblem});
\item we evaluate the performance of the natural analog-digital
architecture in terms of the rate-distortion tradeoff for the binary
tree structure problem (Section~\ref{sec:tree_ach});
\item under the assumption that certain variables have positive
  variance, we derive a novel outer bound to the rate-distortion 
region---this involves a careful use of the entropy-power inequality
(extracting critical ideas from \cite{Oohama05, PTR04}) and is one
of the most important technical contributions of this paper
(Section~\ref{sec:tree_conv});
\item again under the positive variance assumption, we show
  that the outer bound to the rate-distortion region indeed matches the
  inner
bound derived by evaluating the natural analog-digital separation
architecture (Section~\ref{sec:tree_conv});
\item using a continuity argument, we relax the positive variance
  assumption and show that the separation
    architecture is optimal for all binary tree structure
   problems (Section~\ref{sec:treemain});
\item finally, we show that Gaussian sources are the worst case in 
the sense that
a non-Gaussian source has a larger rate-distortion region than
a Gaussian source with the same covariance matrix, so long as the
Gaussian source satisfies the 
tree structure (Section~\ref{sec:tree_robustness}).
\end{itemize}
\subsection{Binary Gauss-Markov Trees}
\label{sec:tree}
\begin{figure}[ht]
\begin{center}
\scalebox{1.2}{\input{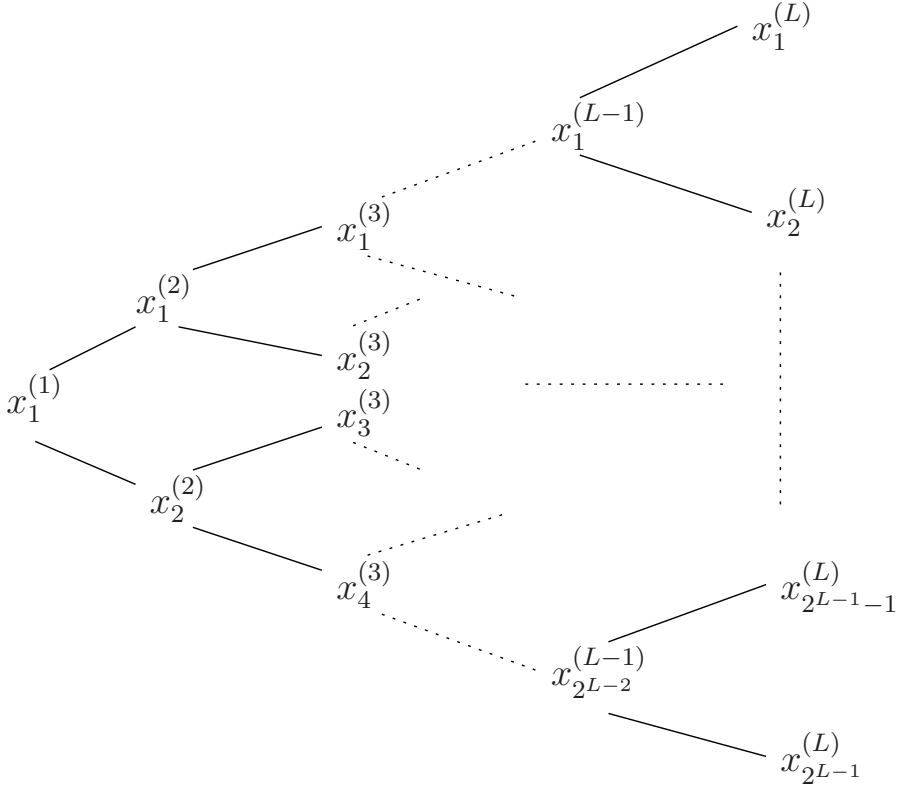}}
\end{center}
\caption{The binary tree structure.} \label{fig:tree}
\end{figure}
Consider the  Markov binary tree structure of Gaussian random
variables depicted in  Figure~\ref{fig:tree}. Formally, the 
Gauss-Markov tree structure represents the following Markov chain
conditions: consider the node denoted by the random variable
$x_i^{(k)}$. We define the set of left descendants, the set of right
descendants, and the
 tree of $x_i^{(k)}$ to be
\begin{eqnarray*}
\mathcal{L}\lp x_i^{(k)}\rp & = & \left\{ x_j^{(l)} : l > k,
\frac{2^{l}(i-1)}{2^k} < j \leq \frac{2^{l}(i-0.5)}{2^k} \right\}, \\
\mathcal{R}\lp x_i^{(k)}\rp & = & \left\{ x_j^{(l)} : l > k,
\frac{2^{l}(i-0.5)}{2^k} < j \leq \frac{2^{l}i}{2^k} \right\},\\
\mathcal{T}\lp x_i^{(k)}\rp & = & \left\{ x_i^{(k)}\right\}\cup
\mathcal{R}\lp x_i^{(k)}\rp \cup \mathcal{L}\lp x_i^{(k)}\rp,
\end{eqnarray*} respectively.
We define the set of nodes $\mathcal{P}\lp x_i^{(k)}\rp$ to be:
\begin{eqnarray*}
\left\{x_j^{(l)}: \forall~j,l \right\} \backslash \;
\mathcal{T}\lp x_i^{(k)}\rp.
\end{eqnarray*}
Then, by definition,
the Markov chain condition given by Figure \ref{fig:tree} says
that  conditioned on the random variable $x_i^{(k)}$,
the sets of random 
variables $\mathcal{P}\lp x_i^{(k)}\rp$, $\mathcal{L}\lp x_i^{(k)}\rp$,
and $\mathcal{R}\lp x_i^{(k)}\rp$ are independent; further,
this is true for all pairs $(i,k)$.

\subsubsection{A Specific Construction}
Now  consider the following  specific construction of $x_i^{(k)}$s
that satisfies the Markov chain structure in Figure \ref{fig:tree}.
Let $m$, $k$, and $i$ denote the time index, the tree depth index, and
the node within the tree depth index, respectively. Then define
\beqa
x_{2i-1}^{(k+1)}(m) & = & \alpha_{2i-1}^{(k+1)}x_{i}^{(k)}(m) + n_{2i-1}^{(k+1)}(m), \label{eq:t1}\\
x_{2i}^{(k+1)}(m) & = & \alpha_{2i}^{(k+1)}x_{i}^{(k)}(m) +
n_{2i}^{(k+1)}(m), \label{eq:t2}
\eeqa
where the indices vary as: \beqa m &=& 1,\ldots,n, \\
k &=& 1,\ldots, L-1 \\ i &=& 1,\ldots,2^{k-1}. \eeqa Here
$\alpha_{2i-1}^{(k+1)}$ and $\alpha_{2i}^{(k+1)}$ are real numbers.
The random variables \beq \lbr n_{i}^{(k)}(m), \quad k=2,\ldots ,L,
\quad i=1,\ldots ,2^{k-1}, \quad m=1,\ldots ,n\rbr\eeq are
independent Gaussian random variables (with zero mean and variance
$\sigma_{n_{i}^{(k)}}^2$ for the index pair $(k,i)$ and any $m$).
Further, these random variables are all independent of the root
random variables
$$\lbr x^{(1)}_1(m),\quad m=1,\ldots ,n\rbr. $$
Finally  let the root random variables
  $$\lbr x^{(1)}_1(m), \quad m=1,\ldots ,n\rbr $$
be a collection of i.i.d.\ Gaussian random
variables with zero mean and variance  $\sigma_{x_1^{(1)}}^2$.
From this construction, it readily follows that the random variables satisfy
the tree structure in Figure~\ref{fig:tree}. Formally:
\begin{claim}
For this construction, the $x_i^{(k)}$ satisfy the Markov chain
conditions in Figure~\ref{fig:tree}.
\end{claim}
\subsubsection{Necessity of Construction}
Conversely, this is also the most general way of
constructing jointly Gaussian random variables that satisfy the binary
tree
 structure. We state this formally below:
\begin{claim}
Any zero-mean, jointly Gaussian $\lbr x_i^{(k)},\quad k=1,\ldots ,L, \quad
i=1,\ldots ,2^{k-1}\rbr $ that satisfy the Markov tree structure in
Figure \ref{fig:tree} can be represented using the above
construction (c.f.\ Equations~\eqref{eq:t1} and~\eqref{eq:t2}).
\end{claim}
{\bf Proof}: The steps are routine: For a fixed $1\leq k < L$ and
$1\leq i\leq 2^{k-1}$, consider the Gaussian random variable
$x_{2i-1}^{(k+1)}$. Since it is jointly Gaussian with all of the
variables in $\C{P}(x_{2i-1}^{(k+1)})$, we can write:
\beq\label{eq:mmse}
 x_{2i-1}^{(k+1)} = \E\Lbr x_{2i-1}^{(k+1)} \mid
\C{P}(x_{2i-1}^{(k+1)})\Rbr + n_{2i-1}^{(k+1)}. \eeq Here the random variable
$n_{2i-1}^{(k+1)}$ is Gaussian and independent of all the nodes
in $\C{P}(x_{2i-1}^{(k+1)})$. Further, the conditional
expectation in Equation~\eqref{eq:mmse} is simply the {\em linear}
conditional expectation that is particularly simple (this is due to
the Markov chain conditions imposed by the tree structure):
specifically, conditioned on $x_{i}^{(k)}$ the random variable of
focus, $x_{2i-1}^{(k+1)}$, is independent of all the other variables
in $\C{P}(x_{2i-1}^{(k+1)})$.
Thus we can write \beq\label{eq:lmmse} \E\Lbr
x_{2i-1}^{(k+1)} \mid \C{P}(x_{2i-1}^{(k+1)})
  \Rbr = \alpha_{2i-1}^{(k+1)} x_i^{(k)}, \eeq
for some real number $\alpha_{2i-1}^{(k+1)}$. Substituting
Equation~\eqref{eq:lmmse} in Equation~\eqref{eq:mmse}, we have
derived Equation~\eqref{eq:t1}. The derivation of
Equation~\eqref{eq:t2} is analogous. Since $n_i^{(k)}$ is independent
of $\C{P}(x_{i}^{(k)})$ for all $i$ and $k$, the required
independence conditions hold and the conclusion follows.
\hfill $\Box$
\subsection{Problem Statement}
\label{sec:binarytreeproblem}
\begin{figure}[ht]
\begin{center}
\scalebox{1.0}{\input{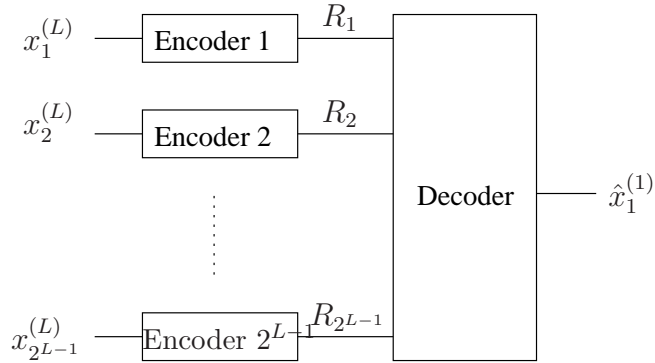}}
\end{center}
\caption{The problem setup.} \label{fig:setup}
\end{figure}

Denote  the vector \beq x_{1,n}^{(1)} \df \lp x_1^{(1)}(1), \ldots,
x_1^{(1)}(n)\rp . \eeq  Similar notation will be used for other
vectors to be introduced later. Consider the following distributed source
coding problem depicted in Figure~\ref{fig:setup}:
 There are $2^{L-1}$ distributed encoders each having access to
 a memoryless observation sequence: encoder $i$ observes the memoryless
 random process $x_{i, n}^{(L)}$. The goal of each encoder is
 to map the observation into a {\em discrete} set (encoder $i$
 maps its length-$n$ observation into a discrete set $C_i$). The
 encoded observation is then conveyed to the central decoder on
 rate-constrained links. The rate of communication from encoder $i$
 to the decoder is
 $$
 \frac{1}{n} \log |C_i|.
 $$
 The decoder forms an estimate $\hat{x}_{1,n}^{(1)}$ of the {\em root} of
the binary tree, $x_{1,n}^{(1)}$, based on the
 messages $C_1,\ldots, C_{2^{L-1}}$. The average distortion
of the reconstruction is
$$
\frac{1}{n}\sum_{m=1}^n \E\left[ \left( x_{1}^{(1)}(m) -
\hat{x}_{1}^{(1)}(m)\right)^2 \right].
$$
The goal  is to characterize the set of achievable rates and distortions
$(R_1,\cdots,R_{2^{L-1}},d)$, i.e., those such that there exists an
encoder and decoder such that
$$
R_i \ge \frac{1}{n} \log |C_i| \quad \text{for all $i$}
$$
and
$$
d \ge \frac{1}{n}\sum_{m=1}^n \E\left[ \left( x_{1}^{(1)}(m) -
\hat{x}_{1}^{(1)}(m)\right)^2 \right].
$$. We denote the closure of this set by $\C{RD}^{*}$.

We note that two special cases of this problem have been resolved in
the literature:
\begin{itemize}
\item $L=1$ is the single-user Gaussian source coding problem with quadratic distortion,
\item $L=2$ is the Gaussian CEO problem solved in \cite{Oohama05, PTR04}.
\end{itemize}
The recent work in  \cite{Oohama06} studies a special case of the
general tree structure depicted in
Figure~\ref{fig:tree}.\footnote{As an aside, we note that the
material in  \cite{Oohama06} along with our own previous work
\cite{WTV06} provided the impetus to the present work.} While a
general outer bound is derived in \cite{Oohama06} for that special
case of the tree structure, it is shown to be tight only for a
certain range of the parameters in the problem (the distortion
constraint and the covariance matrix of the Gaussian sources).

Our main result is that a natural strategy of point-to-point Gaussian
vector quantization followed by Slepian-Wolf binning
is optimal for any $L$. In the next section we formally
present the natural achievable strategy and then state our main
result. In the subsequent section, we prove a novel outer bound
and use it to establish the main result.

\subsection{Analog-Digital Separation Strategy}\label{sec:tree_ach}

\begin{figure}[ht]
\begin{center}
\scalebox{.95}{\input{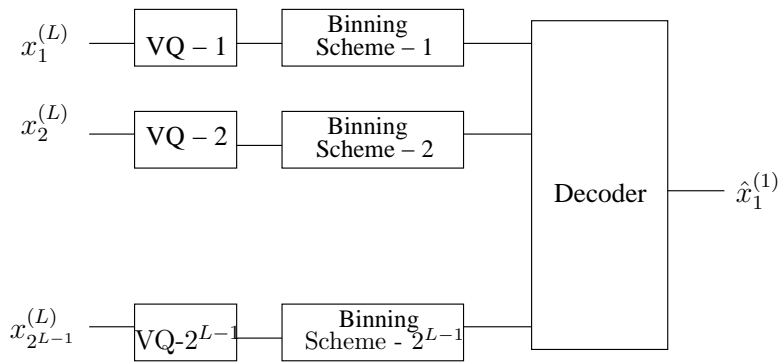}}
\end{center}
\caption{The natural separation scheme.} \label{fig:natsep}
\end{figure}

The  natural achievable analog-digital separation strategy is
depicted in Figure \ref{fig:natsep}: each encoder  first vector
quantizes the observation  as in point-to-point Gaussian rate distortion
theory, and then codes the quantizer outputs using a Slepian-Wolf
binning scheme.
The rate tuples needed by this architecture to satisfy the
distortion constraint can be calculated by the so-called Berger-Tung
inner bound \cite{B78}: let
\beq
 \ve{u} \df \lp u_1, u_2, \cdots, u_{2^{L-1}}\rp
\eeq
denote a vector of $2^{L-1}$ jointly Gaussian random variables. Consider
the set $\C{U}(d)$ of $\ve{u}$ such that
\begin{itemize}
\item For each $i = 1,\ldots,2^{L-1}$, $u_i$ satisfies
  \beq u_i  =  \alpha_i x_i^{(L)} + w_i,\eeq
  where $\alpha_1,\ldots,\alpha_{2^{L-1}}$ are constants and
  $w_1,\ldots ,w_{2^{L-1}}$ are independent zero-mean Gaussian random
  variables that are also independent of the $x_i^{(k)}$s. It is
  convenient to assume
  that $\alpha_i \in [0,1]$ and that
  $w_i$ has variance $(1 - \alpha_i^2)\sigma^2_{x_i^{(L)}}$,
  so that $x_i^{(L)}$ and $u_i$ have the same variance. This assumption
  incurs no loss of generality.
\item $\ve{u}$ satisfies
\beq \E\left[\left(x_1^{(1)} -\E[x_1^{(1)}|\ve{u}]\right)^2\right]
\leq d.\eeq
\end{itemize}
Now, consider
\beq
\C{A} \subseteq \lbr 1,\ldots ,2^{L-1}\rbr.
\eeq
Denote the set
\beq \left\{ u_i: i\in\C{A}\right\}
\df \ve{u}_{\C{A}}.
\eeq
Similar notation will be used for other vectors introduced later. We now have:
\begin{lemma}\label{lemma:BT}[Berger-Tung inner bound \cite{B78}]
The analog-digital separation architecture achieves convex hull
of the rate-distortion
 region
\begin{multline}
\label{eq:inner}
\C{RD}_{\rm in} \df \Big[\lp R_1,\cdots, R_{2^{L-1}},d \rp :
\; \exists \; \ve{u} \in \C{U}(d)\ni
\forall\C{A}\subseteq\lbr 1,\ldots ,2^{L-1}\rbr,  \\
 \sum_{i \in \C{A}} R_i \geq I\lp\ve{x}_{\C{A}}^{(L)};
\ve{u}_{\C{A}} | \ve{u}_{\C{A}^c} \rp\Big].
\end{multline}
In particular, $\C{RD}^*$ contains $\Co(\C{RD}_{\rm{in}})$, where
$\Co(\cdot)$ denotes the closure of the convex hull.
\end{lemma}
The region $\C{RD}_{\rm{in}}$ can be explicitly computed for a
given covariance matrix for the observed Gaussian
sources. This computation is aided by the following
combinatorial structure of
the set $\C{RD}_{\rm{in}}$.

\subsubsection{Combinatorial Structure of $\C{RD}_{\rm in}$}
Consider a specific $\ve{u} \in \C{U}(d)$ (this parameterizes a
specific choice of the analog-digital separation architecture) and
the rate tuples $\lp R_1,\ldots ,R_{2^{L-1}}\rp$ that satisfy the
conditions
\beq\label{eq:achieve}
\sum_{i\in\C{A}} R_i \geq
f\lp\C{A}\rp, \quad \forall \C{A} \subseteq \lbr 1,\ldots
,2^{L-1}\rbr
\eeq
where
\beq
f\lp\C{A}\rp \df I\lp \ve{x}_{\C{A}}^{(L)} ; \ve{u}_{\C{A}} |
\ve{u}_{\C{A}^c}\rp. \eeq Consider the following properties of the
set function $f$ for all $\C{A}_1,\C{A}_2\subseteq \lbr 1,\ldots
,2^{L-1}\rbr$. We have $f\lp\phi\rp \df 0$.
\begin{lemma}
\beqa
f(\C{A}_1) & \geq &  0,\label{eq:contrapoly1} \\
 f\lp \C{A}_1 \cup \lbr t\rbr\rp & \geq & f\lp \C{A}_1\rp ,\; \forall t \in
 \lbr 1,\ldots ,2^{L-1}\rbr , \label{eq:contrapoly2} \\
f\lp \C{A}_1\cup \C{A}_2\rp + f\lp \C{A}_1\cap \C{A}_2\rp & \geq
 & f\lp  \C{A}_1\rp + f\lp \C{A}_2\rp .\label{eq:contrapoly3}
\eeqa \label{prop:contrapoly}
\end{lemma}
\prf Equation~\eqref{eq:contrapoly1} follows from the non-negativity of
mutual information.
Equation~\eqref{eq:contrapoly2} follows from the chain rule of
mutual information: for $t\notin \C{A}_1$, we have
\beqan
f\lp \C{A}_1\cup\lbr t\rbr\rp
& = & I\lp \ve{u}_{\C{A}_1}\; ;\; x^{(L)}_t \ve{x}^{(L)}_{\C{A}_1}\mid
\ve{u}_{\C{A}_1^c}\rp + 
  I\lp u_t\; ; \; x^{(L)}_t \ve{x}^{(L)}_{\C{A}_1} \mid 
   \ve{u}_{(\C{A}_1 \cup \{t\})^c}
\rp ,\\
& \geq & I\lp \ve{u}_{\C{A}_1}\; ;\; x^{(L)}_t \ve{x}^{(L)}_{\C{A}_1}
    \mid \ve{u}_{\C{A}_1^c}\rp
 ,\\
& = & f\lp \C{A}_1\rp . \eeqan 
Finally, consider~(\ref{eq:contrapoly3}). Let
$$
B = \{i: \Var(u_i|x_i^{(L)}) > 0\}.
$$
Suppose $i \in (\C{A}_1 \cup \C{A}_2) \cap B^c$. If $\Var(x_i^{(L)}|
  \ve{x}^{(L)}_{(\C{A}_1 \cup \C{A}_2)^c}) > 0$, then
  $f(\C{A}_1 \cup \C{A}_2) = \infty$, so \eqref{eq:contrapoly3} 
  trivially holds. 
  If $\Var(x_i^{(L)}|x^{(L)}_{(\C{A}_1 \cup \C{A}_2)^c}) = 0$, then
$$
\Var(x^{(L)}_i|\ve{x}^{(L)}_{\C{A}_1^c}) = 
  \Var(x_i^{(L)}|\ve{x}^{(L)}_{\C{A}_2^c}) =
  \Var(x^{(L)}_i|\ve{x}^{(L)}_{\C{A}_1^c \cup \C{A}_2^c}) = 0,
$$
so $\C{A}_1$ and $\C{A}_2$ can be replaced with $\C{A}_1 \backslash
  \{i\}$ and $\C{A}_2 \backslash \{i\}$, respectively, without
  affecting the validity of~(\ref{eq:contrapoly3}). By 
  repeating this process as many times as necessary, we may assume that
   $\C{A}_1 \cup \C{A}_2 \subset B$.

  This case requires the use of the Markov property
satisfied by  $\ve{u}$: in particular, we have by construction
$$
\ve{u}_{\C{A}_1} \markov \ve{x}^{(L)}_{\C{A}_1} \markov 
   \ve{u}_{\C{A}_1^c},
$$
meaning that these tree variables form a Markov chain in the
specified order. Thus we can write
\beq h\lp \ve{u}_{\C{A}_1}\mid \ve{x}^{(L)}_{\C{A}_1},\ve{u}_{\C{A}_1^c}
    \rp = \sum_{i\in\C{A}_1} h\lp u_i \mid
x_i^{(L)} \rp .\label{eq:markovuse} \eeq 
Now we rewrite $f(\C{A}_1)$
as, using \eqref{eq:markovuse},
 \beqa f\lp \C{A}_1\rp & = & h\lp
\ve{u}_{\C{A}_1}\mid \ve{u}_{\C{A}_1^c}\rp - \sum_{i\in
  \C{A}_1} h\lp u_i\mid x_i^{(L)}\rp \label{eq:markovuse1} \\
 & = & h\lp \ve{u}\rp - h\lp  \ve{u}_{\C{A}_1^c}\rp - \sum_{i\in
  \C{A}_1} h\lp u_i\mid x_i^{(L)}\rp.\label{eq:markovuse2}
\eeqa
It follows from  \eqref{eq:markovuse2} that we have shown
\eqref{eq:contrapoly3} if
\begin{align*}
 h\lp \ve{u}_{\C{A}_1^c}\rp + h\lp \ve{u}_{\C{A}_2^c}\rp & \geq h\lp
 \ve{u}_{(\C{A}_1\cup \C{A}_2)^c} 
   \rp + h\lp \ve{u}_{(\C{A}_1 \cap \C{A}_2)^c}\rp ,
\intertext{i.e.,}
 h\lp \ve{u}_{\C{A}_1^c-\C{A}_2^c} \mid \ve{u}_{\C{A}_1^c\cap \C{A}_2^c}\rp
 & \geq  h\lp \ve{u}_{\C{A}_1^c-\C{A}_2^c}\mid \ve{u}_{\C{A}_2^c}\rp,
\end{align*}
which is true
since conditioning cannot increase the differential entropy. \hfill $\Box$

A polyhedron such as the one in \eqref{eq:achieve} with the {\em
rank} function $f$ satisfying the properties in
Lemma~\ref{prop:contrapoly} is called a {\em contra-polymatroid}. A
generic reference to the class of polyhedrons called {\em matroids}
is \cite{Matroid} and  applications to information theory are in
\cite{TH97} where natural achievable regions of the  multiple access
channel are shown to be {\em polymatroids} and in \cite{CZBW04,WV06} where
natural achievable regions are shown to be contrapolymatroids. An
important property of contra-polymatroids is summarized in Lemma~3.3
of \cite{TH97}: the characterization of its vertices. For $\pi$ a
permutation on the set $\lbr 1,\ldots, 2^{L-1}\rbr$, let
\[ b^{\lp \pi\rp}_{\pi_i}  \df  f\lp \lbr
\pi_1,\pi_2,\ldots ,\pi_i\rbr\rp - f\lp\lbr \pi_1,\pi_2,\ldots
,\pi_{i-1}\rbr\rp,\; i=1\ldots 2^{L-1},\] and $\ve{b}^{\lp \pi\rp} = \lp
b^{\lp\pi\rp}_{\pi_1},\ldots ,b^{\lp\pi\rp}_{\pi_{2^{L-1}}}\rp$.
Then the $2^{L-1}!$
points $\lbr \ve{b}^{\lp\pi\rp},\;\pi \mbox{ a permutation}\rbr$,
are the vertices of (and hence belong to) the contra-polymatroid
\eqref{eq:achieve}. We use this result to conclude that all of the
constraints in \eqref{eq:achieve} are tight for some rate tuple
and there is a
computationally simple way to find the vertex that
leads to a minimal linear functional of
the rates~\cite{TH97}.

\subsection{An outer bound for a special case}\label{sec:tree_conv}

We first focus on the case in which $\sigma^2_{n_i^{(k)}} > 0$ for
all $i$ and $k$. We abbreviate this condition by saying that ``all
of the noise variances are positive.''
To derive our outer bound, we need the following definitions:
\begin{itemize}
\item
Fix $1 \le k \le L-1$ and $1\leq i\leq 2^{k-1}$ and define the function \beq
\label{eq:fxik_defn} f_{x_i^{(k)}}\lp r_1,r_2\rp \df
\frac{1}{2}\log\lp 1 +
\frac{\alpha_{2i-1}^{(k+1)}
  \sigma^2_{n_i^{(k)}}}{\sigma^2_{n_{2i-1}^{(k+1)}}} \lp 1 -
   e^{-2r_1}\rp + \frac{\alpha_{2i}^{(k+1)}
   \sigma^2_{n_i^{(k)}}}{\sigma^2_{n_{2i}^{(k+1)}}} \lp
1 - e^{-2r_2}\rp \rp, \quad r_1,r_2\geq 0. \eeq
\item
For node $x_i^{(k)}$, we define the set of associated {\em
observations} to be
\begin{eqnarray}
\C{O}\lp x_i^{(k)}\rp & = & \left\{j: \frac{2^{L}(i-1)}{2^{k}} < j
\leq \frac{2^{L}i}{2^{k}} \right\}.\label{eq:observe}
\end{eqnarray}
\item
To each node in the binary tree structure of Figure~\ref{fig:tree} we
associate a nonnegative number, known as \emph{noise-quantization rate}.
Specifically
associate $r_i^{(k)}$ with the node $x_i^{(k)}$. 
A physical interpretation for
the nomenclature ``noise quantization rate'' will be available during the 
proof of the outer bound.
\item For each node $x_i^{(k)}$ define the set $\ve{r}_{\C{A},\C{A}^c}
  (x_i^{(k)})$ to be
the set of noise-quantization rates (say, $r_j^{(l)}$) of the
variables (say $x_j^{(l)}$) in the tree of $x_i^{(k)}$ whose associated
observations are entirely in $\C{A}$ or $\C{A}^c$ and are such 
that none of the ancestors of $x_j^{(l)}$
have this property. Formally,
\begin{eqnarray*}
\ve{r}_{\C{A},\C{A}^c}\lp x_i^{(k)}\rp & = & \left\{r_j^{(l)}:~x_j^{(l)} \in
\C{T}\lp x_i^{(k)}\rp,~ \C{O}\lp x_j^{(l)}\rp \subset \C{A} \ \mbox{or} \
   \C{O}\lp x_j^{(l)}\rp \subset \C{A}^c \right.
\\ \nonumber &&  \not\exists~ x_a^{(b)}~\in \C{T}\lp x_i^{(k)}\rp
~\mbox{with}~\C{O}\lp x_a^{(b)}\rp \subset \C{A}~\mbox{or} \
   \C{O}\lp x_a^{(b)}\rp \subset \C{A}^c,\\
&&\left.\mbox{and}~x_j^{(l)}~\in~\C{R}\lp x_a^{(b)}\rp\cup\C{L}(x_a^{(b)})
\right\}.
\end{eqnarray*}
Likewise, we let  $\ve{r}_{\C{A}} (x_i^{(k)})$  denote
the set of noise-quantization rates of variables in the  tree
of $x_i^{(k)}$ whose associated
observations are entirely in $\C{A}$ and are such that none of the
ancestors have this property. Formally,
\begin{eqnarray}\nonumber
\ve{r}_{\C{A}}\lp x_i^{(k)}\rp & = & \left\{r_j^{(l)}:~x_j^{(l)} \in
\C{T}\lp x_i^{(k)}\rp,~ \C{O}\lp x_j^{(l)}\rp \subset \C{A} \, \right.
\\ \nonumber &&  \not\exists~ x_a^{(b)}~\in \C{T}\lp x_i^{(k)}\rp
~\mbox{with}~\C{O}\lp x_a^{(b)}\rp \subset \C{A},\\\label{eq:raxik}
&&\left.\mbox{and}~x_j^{(l)}~\in~\C{R}\lp x_a^{(b)}\rp\cup\C{L}(x_a^{(b)})
\right\}.
\end{eqnarray}

\item Define the following set of noise-quantization rates
 $\lp r_i^{(k)}, 1\leq k\leq L, 1\leq i\leq 2^{k-1}\rp$:
\beqa
\C{F}_r(d) & = & \left\{r_i^{(k)} \geq 0,
 r_1^{(1)} \geq \frac{1}{2}\log\frac{\sigma_{x_1^{(1)}}^2}{d},
 r_{i}^{(k)} \leq f_{x_{i}^{(k)}}\lp r_{2i-1}^{(k+1)},r_{2i}^{(k+1)}\rp
 \right\}. \label{eq:frd}
\eeqa

\item We next implicitly define a collection of functions of the
  noise-quantization rates.
  Consider a set of noise-quantization rates
$\lp r_i^{(k)}, 1\leq k\leq L, 1\leq i\leq 2^{k-1}\rp$ in $\C{F}_r(d)$.
Then for any $i$ and $k$, we have
$$
r_{i}^{(k)} \leq f_{x_{i}^{(k)}}\lp r_{2i-1}^{(k+1)},r_{2i}^{(k+1)}\rp.
$$
Since $f_{x_i^{(k)}}$ is increasing in both arguments, this
implies
\begin{align*}
r_{i}^{(k)} & \leq f_{x_{i}^{(k)}}\lp r_{2i-1}^{(k+1)},
   f_{x_{2i}^{(k+1)}}\lp r_{4i-1}^{(k+2)},r_{4i}^{(k+2)}\rp\rp \\
r_{i}^{(k)} & \leq f_{x_{i}^{(k)}}\lp 
      f_{x_{2i-1}^{(k+1)}}\lp r_{4i-3}^{(k+2)}, r_{4i-2}^{(k+2)} \rp,
      r_{2i}^{(k+1)}\rp \\
r_{i}^{(k)} & \leq f_{x_{i}^{(k)}}\lp 
  f_{x_{2i-1}^{(k+1)}}\lp r_{4i-3}^{(k+2)}, r_{4i-2}^{(k+2)}\rp,
    f_{x_{2i}^{(k+1)}}\lp r_{4i-1}^{(k+2)},r_{4i}^{(k+2)}\rp \rp.
\end{align*}
By repeating this substitution process, we may obtain an upper bound
on $r_i^{(k)}$ in terms of the noise-quantization rates in
$\ve{r}_{\C{A},\C{A}^c}\lp x_i^{(k)}\rp$. We implicitly
define 
\beq
\label{eq:telescope}
f_{x_i^{(k)}}^{\C{A},\C{A}^c} \lp \ve{r}_{\C{A},\C{A}^c}\lp x_i^{(k)}\rp \rp
\eeq
to be this upper bound. (By convention, if
$$
\ve{r}_{\C{A},\C{A}^c}(x_i^{(k)}) = \{r_i^{(k)}\},
$$ 
then we define this upper bound to be $r_i^{(k)}$ itself.)
We then let
$$
f_{x_i^{(k)}}^\C{A}\lp \ve{r}_{\C{A}}\lp x_i^{(k)}\rp \rp
$$
denote the function of $\ve{r}_{\C{A}}\lp x_i^{(k)}\rp$ obtained
by evaluating the function in~(\ref{eq:telescope}) 
with all of the noise quantization rates in
$$
\ve{r}_{\C{A},\C{A}^c}\lp x_i^{(k)}\rp
\backslash \ \ve{r}_{\C{A}}\lp x_i^{(k)}\rp
$$  set equal to zero.
The significance of this function will be apparent in the
proof of the outer bound.

\item For any set
\beq
\C{A} \subseteq \lbr 1, 2,\ldots, 2^{L-1}\rbr,
\eeq
we define the {\em ancestors set at level $k$} to be
\begin{eqnarray}\label{eq:ancestors}
\C{A}^{(k)}\df \left\{ i : \C{O}(x_i^{(k)}) \cap \C{A} \not= \Phi  \right\},
\end{eqnarray}
where $\Phi$ denotes the empty set.
\end{itemize}

Consider the following region, $\C{RD}_{\rm out}$, defined as
\begin{multline}
\C{RD}_{\rm out} =
\Big\{ \lp R_1,\cdots, R_{2^{L-1}},d\rp : \exists \left\{ r_i^{(k)}
\right\} \in \C{F}_r(d) \ni  \\
  \forall \C{A}
\subseteq \lbr 1,\ldots ,2^{L-1}\rbr
\sum_{i \in \C{A}} R_i \geq \sum_{k=1}^{L}\sum_{i \in
\C{A}^{(k)}} \left(r_i^{(k)}
-f_{x_{i}^{(k)}}^{\C{A}^c}\left(\ve{r}_{\C{A}^c}(x_i^{(k)})\right)\right)
\Big\}.\label{eq:outer}
\end{multline}

This constitutes an outer bound to the rate-distortion region
of the binary tree structure problem:
\begin{lemma}\label{lemma:outer}
For the binary tree structure problem in which all of the noise
variances are positive,
\beq
\C{RD}^* \subset \C{RD}_{\rm out}.
\eeq
\end{lemma}
{\bf Proof}: See Appendix~\ref{app:outer}.

We next show that the outer bound just derived matches
the inner bound derived from the analog-digital separation
architecture (c.f.\ Lemma~\ref{lemma:BT}).
 Recall that we 
use $\Co(\cdot)$
to denote the closure of the convex hull of a given set.
\begin{lemma}\label{lemma:matchup}
For the binary tree structure problem in which all of the noise
variances are positive,
\beq
\C{RD}_{\rm out} = \Co\lp\C{RD}_{\rm in}\rp.
\eeq
\end{lemma}
{\bf Proof}: See Appendix~\ref{app:matchup}.

\subsection{Main Result}\label{sec:treemain}

Using a continuity argument, one can relax the assumption that
all of the noise variables have positive variance.
This allows us to
conclude our first main result of this paper:  the optimality of
the analog-digital separation architecture in achieving the rate-distortion
region of the binary tree structure problem.
\begin{theorem}\label{thm:main}
For the binary tree structure problem, the optimal rate-distortion
region
is achieved by the analog-digital separation architecture,
$$
\C{RD}^* = \Co\lp\C{RD}_{\rm{in}}\rp.
$$
\end{theorem}

\prf
See Appendix~\ref{appendix:cont}.
\hfill $\Box$

\subsection{Worst-Case Property}\label{sec:tree_robustness}

Up to this point we have assumed that
the source variables are jointly Gaussian. In this section,
we justify this assumption by showing that the rate-distortion region
for other distributions with the same covariance are only larger.

Let $\lp x_i^{(k)} \rp$ be a Gaussian source satisfying the tree
structure as before. 
Let $$\lp \tilde{x}_1^{(1)}, \tilde{x}_1^{(L)}, \ldots, 
\tilde{x}_{2^{L - 1}}^{(L)} \rp$$ be an alternate source
with the same covariance of 
$$\lp x_1^{(1)}, x_1^{(L)}, \ldots, x_{2^{L - 1}}^{(L)} \rp.$$
Note that the alternate source need not be part of a Markov
tree. Let $\widetilde{\C{RD}}^*$ denote the rate-distortion
region of the alternate source.

The separation-based architecture yields an inner bound
on the rate-distortion region of the alternate source. 
Specifically, let
$\widetilde{\C{RD}}_{in}$ denote the region obtained by replacing
$\lp x_1^{(1)}, x_1^{(L)}, \ldots, 
x_{2^{L - 1}}^{(L)} \rp$ with
$\lp \tilde{x}_1^{(1)}, \tilde{x}_1^{(L)}, \ldots, 
\tilde{x}_{2^{L - 1}}^{(L)} \rp$ 
in the discussion in Section~\ref{sec:tree_ach}. Then
$$
\Co\lp\widetilde{\C{RD}}_\mathrm{in}\rp \subset
   \widetilde{\C{RD}}^*.
$$

\begin{theorem}
\label{theorem:robust}
A Gaussian source satifying the binary tree
structure has the smallest rate-distortion region
for its covariance:
$$
\C{RD}^* \subset \widetilde{\C{RD}}^*.
$$
In fact, the separation-based architecture has the
most difficulty compressing a Gaussian source in the 
sense that
\beq
\label{eq:robust2}
\C{RD}^* = \Co\lp\C{RD}_{\mathrm{in}} \rp
  \subset \Co\lp\widetilde{\C{RD}}_{\mathrm{in}}\rp \subset 
    \widetilde{\C{RD}}^*.
\eeq
\end{theorem}

\prf
See Appendix~\ref{app:robust}.
\hfill $\Box$

\section{Tree Structure and the Many-Help-One Problem}
\label{sec:manyhelpone}
\begin{figure}[ht]
\begin{center}
\scalebox{.95}{\input{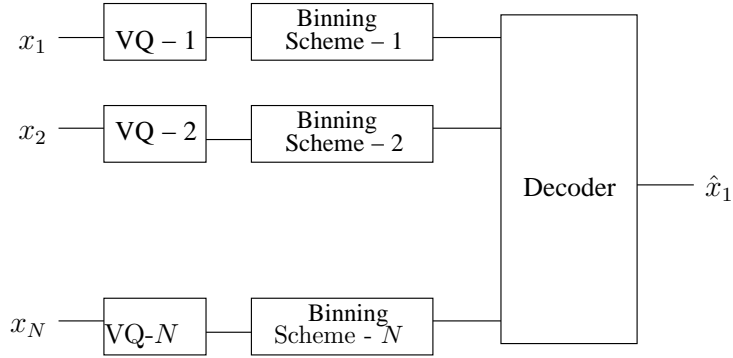}}
\end{center}
\caption{The natural analog-digital separation architecture.}
\label{fig:natsep_manyhelpone}
\end{figure}

We now turn to the main problem 
of interest: the many-help-one distributed source
coding problem. As in the tree structure problem, there is a natural analog-digital
separation architecture that is a candidate solution. This is illustrated in
Figure~\ref{fig:natsep_manyhelpone}.

\subsection{Main Result}
Our main result is a sufficient condition under which the
analog-digital separation architecture is optimal. To 
state it, we first define a {\em general} Gauss-Markov tree: it is
made up of jointly Gaussian random variables and respects the Markov
conditions implied by the tree structure. The only extra feature
compared to the {\em binary} Gauss-Markov tree (c.f.\
Figure~\ref{fig:tree}) is that each node can have any number of
descendants (not just two).

\begin{theorem}\label{thm:main_manyhelpone}
Consider the many-help-one distributed source coding problem
illustrated in Figure~\ref{fig:manyhelpone}. Suppose the observations
$x_1,\ldots, x_N$ can be embedded in a general
Gauss-Markov tree of size $M \geq N$. Then the natural analog-digital separation
architecture (c.f.\ Figure~\ref{fig:natsep_manyhelpone}) achieves
the entire rate-distortion region.
\end{theorem}
{\bf Proof}: The proof is elementary and builds heavily on
Theorem~\ref{thm:main}. We outline the steps below:
\begin{itemize}
\item  A general Gauss-Markov tree can be 
recast as a (potentially larger) binary Gauss-Markov tree with the root being 
identified with any specified node in the original tree.  To see this, we only need to observe
that the Markov chain relations are the same no matter which node is identified as the root.  
\item Next, by potentially increasing the height of the binary tree 
(to $\tilde{L} \geq L$) 
we can  ensure that  the observations $x_1,\ldots ,x_N$ are a subset of the 
$2^{\tilde{L}-1}$ leaves of the binary Gauss-Markov tree. 
If one observation of interest, say $x_i$, is an 
intermediate node of the binary Gauss-Markov
tree we can  effectively make it a 
leaf by adding descendants that are identical (almost surely) 
to $x_i$. 
\end{itemize}
This allows us to convert the many-help-one problem into a binary tree structure problem (with 
potentially more observations than we started out with). The analog-digital 
separation architecture is optimal for this problem (c.f.\ Theorem~\ref{thm:main}). By restricting
the corresponding rate-distortion region to the instance when the rates of the encoders corresponding
to the observations that were not part of the original $N$ are zero, 
we still have the optimality of
the analog-digital separation architecture. This latter rate-distortion region simply corresponds
to the many-help-one problem studied in Figure~\ref{fig:manyhelpone}. This completes the proof.  \hfill $\Box$

\begin{figure}[ht]
\begin{center}
\scalebox{1.2}{\input{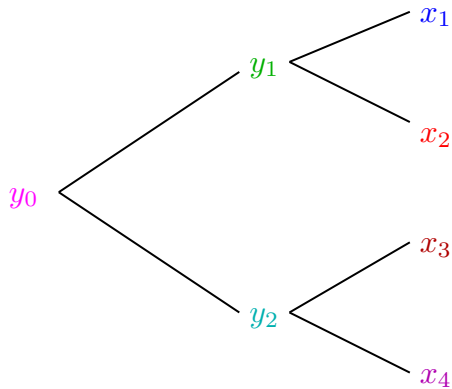}}
\end{center}
\caption{Four observations are embedded in a (binary) Gauss-Markov tree.}
\label{fig:embed}
\end{figure}

\begin{figure}[ht]
\begin{center}
\scalebox{.8}{\input{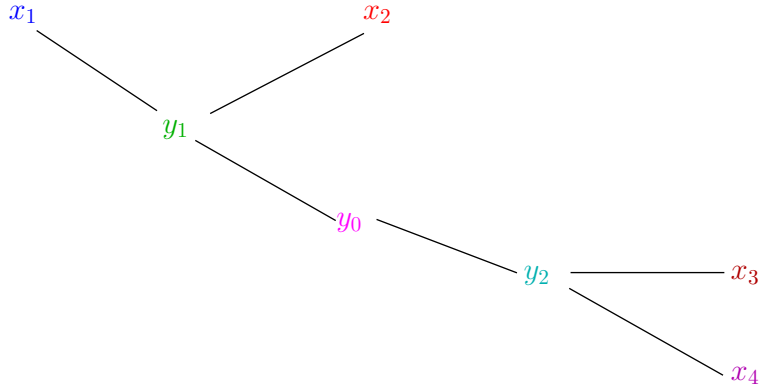}}
\end{center}
\caption{The tree rewritten with $x_1$ as the root.}
\label{fig:re_write}
\end{figure}

\begin{figure}[ht]
\begin{center}
\scalebox{.6}{\input{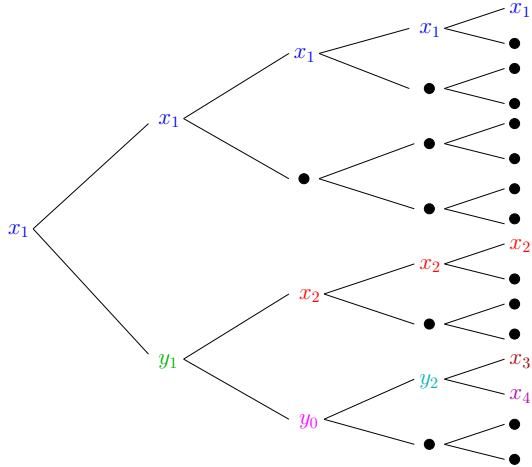}}
\end{center}
\caption{The many-help-one problem rewritten as a binary tree structure problem.}
\label{fig:dummynodes}
\end{figure}

We illustrate the two key steps outlined above with an example with $N=4$. 
Suppose that $x_1,\ldots,x_4$ can be embedded in the tree depicted in
Figure~\ref{fig:embed}. This tree happens to be binary, but unfortunately
the root is not the source of interest, $x_1$.
Figure~\ref{fig:re_write} shows how to construct a new Gauss-Markov tree that still
preserves the Markov conditions but has $x_1$ as its root.  
Finally, a binary Gauss-Markov tree of height 5 is constructed 
that has the original four observations as a subset of its 16 leaf nodes; this is done in 
Figure~\ref{fig:dummynodes}---here any node indicated by a dot 
is simply identically equal (almost
 surely) to its parent node.  Finally we can set 
to zero the rates of all the encoders except those
numbered 1, 9, 13 and  14. This allows us to capture the rate-distortion region of the original
three-help-one problem.

\subsection{Worst-Case Property}

As with our earlier result for the
binary tree structure problem, the Gaussian assumption in 
Theorem~\ref{thm:main_manyhelpone}
can be justified on the grounds that it is the worst-case distribution.
Specifically, as in Section~\ref{sec:tree_robustness}, 
let $\tilde{x}_1,\ldots,\tilde{x}_N$
denote an alternate source with the same covariances as
$x_1,\ldots,x_N$. Let $\widetilde{\C{RD}}^*$ denote the rate-distortion 
region of the source, and let $\widetilde{\C{RD}}_\mathrm{in}$ denote the
inner bound obtained by replacing the source variables in the discussion
in Section~\ref{sec:tree_ach} with the alternate source 
$\tilde{x}_1,\ldots,\tilde{x}_N$.

\begin{theorem}
\label{theorem:robust2}
A Gaussian source that can be embedded in a Gauss-Markov
tree has the smallest rate-distortion region
for its covariance:
$$
\C{RD}^* \subset \widetilde{\C{RD}}^*.
$$
In fact, the separation-based architecture has the
most difficulty compressing a Gaussian source in the
sense that
$$
\C{RD}^* = \Co\lp\C{RD}_{\mathrm{in}}\rp
  \subset \Co\lp\widetilde{\C{RD}}_{\mathrm{in}}\rp \subset 
     \widetilde{\C{RD}}^*.
$$
\end{theorem}

The proof of Theorem~\ref{theorem:robust} applies verbatim here.

\subsection{Tree Structure Condition and Computational Verification}
If $N=2$, then $x_1$ and $x_2$ can always be placed in the trivial
Gauss-Markov tree consisting of these two variables; no embedding is
needed in this case. We note that $N = 2$ corresponds to the
``one-help-one'' problem, whose rate-distortion region has been determined by
Oohama~\cite{Oohama97}. With $N \ge 3$, embedding is not always
possible.  We see an example of this next, where we also see
a simple test for when $N$ linearly
independent variables can themselves be arranged in a tree, without
adding additional variables. We then derive a condition
on the covariance matrix of $x_1,\ldots ,x_N$ that is necessary
for these variables to be  embedded as
the nodes of a general Gauss-Markov tree. Finally, we show that this condition is 
also sufficient when $N=3$.

\subsubsection{Trees Without Embedding}
We next demonstrate a simple test for when $N$ linearly
independent, jointly Gaussian random variables can themselves be arranged in a tree, without
adding additional variables. Without loss of generality, we may assume that
$x_1,\ldots ,x_N$ each has unit variance 
(this can be ensured by normalizing each observation). 
We shall write
$$
\rho_{ij} = \E[x_i x_j].
$$
Suppose that $x_1,\ldots, x_N$ are linearly independent,
and let $\m{K}_x$ denote their (invertible)
covariance matrix.
We will use 
the following fact from the literature (Speed and Kiiveri~\cite{SK86}):
\begin{quote}
 $x_1,\ldots, x_N$ are Markov with respect to a simple,
undirected graph $G$
if and only if for all $i \ne j$ such that $(i,j)$ is not an edge in $G$,
the $(i,j)$ entry of $\m{K}_x^{-1}$ is zero.
\end{quote}
Now let $G$ denote the simple, undirected graph with $x_1,\ldots, x_N$
as the nodes obtained by interpreting $\m{K}_x^{-1} - \m{I}$
as the adjacency
matrix: there is an edge between $x_i$ and $x_j$ if and only if the
$(i,j)$ element of $\m{K}_x^{-1} - \m{I}$ is nonzero.  It follows that
$x_1,\ldots, x_N$ can be arranged in a Gauss-Markov tree if and only
if $G$ is a tree, or more generally, a forest (i.e, a collection of
unconnected trees).

This fact can be illustrated with the following example.
Suppose that $N = 3$ and
\beq
\m{K}_x = \left[
\begin{array}{ccc}
1 & 1/4 & 1/4 \\
1/4 & 1 & 1/4 \\
1/4 & 1/4 & 1
\end{array} \right] .
\eeq
Then
$$
\m{K}_x^{-1} = \frac{1}{9} \left[
\begin{array}{ccc}
   10  & -2 & -2 \\
  -2  & 10 & -2 \\
  -2 &  -2 & 10
\end{array} \right],
$$
which yields a fully-connected graph. Hence $x_1$, $x_2$, and $x_3$
cannot be arranged in a Gauss-Markov tree.

\begin{figure}[ht]
\begin{center}
\scalebox{1.5}{\input{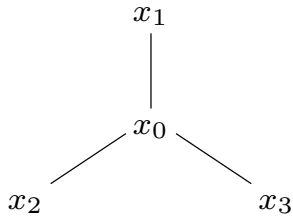}}
\end{center}
\caption{Tree embedding for $x_1$, $x_2$, and $x_3$.}
\label{xtree}
\end{figure}

Nevertheless, it is possible that $x_1,x_2,x_3$ can be \emph{embedded} in
a larger Gauss-Markov tree. Indeed, in this case 
it turns out that it is possible to embed the variables in a tree
of size 4. We offer the following specific construction to demonstrate this
fact. Let $x_0$ be a standard Normal random variable
and let
\begin{align*}
x_1 & = \frac{1}{2} \cdot x_0 + z_1 \\
x_2 & = \frac{1}{2} \cdot x_0 + z_2 \\
x_3 & = \frac{1}{2} \cdot x_0 + z_3
\end{align*}
where $z_1$, $z_2$, and $z_3$ are i.i.d.\ Gaussian with variance $3/4$,
and are independent of $x_0$. The covariance matrix for this quadruple
of variables is
$$
\m{K}_x = \left[
\begin{array}{cccc}
1 & 1/2 & 1/2 & 1/2 \\
1/2 & 1 & 1/4 & 1/4 \\
1/2 & 1/4 & 1 & 1/4 \\
1/2 & 1/4 & 1/4 & 1
\end{array} \right].
$$
The inverse of this matrix is
$$
\m{K}_x^{-1} = \frac{2}{3} \left[
\begin{array}{cccc}
3 & -1 & -1 & -1 \\
-1 & 2 & 0 & 0  \\
-1 & 0 & 2 & 0  \\
-1 & 0 & 0 & 2
\end{array} \right].
$$
with the resulting $G$ being the tree depicted in Fig.~\ref{xtree}.

\subsubsection{Necessary Condition for Tree Embedding}
Even allowing additional variables in the Gauss-Markov tree, it can turn out
that embedding is impossible. Towards understanding the situation better, 
we derive a necessary
condition for $x_1,\ldots,x_N$ to be embeddable. It turns out 
that this condition is 
also  sufficient when $N = 3$.
\begin{prop}
\label{embed}
Let $N \ge 3$.
If $x_1,\ldots,x_N$ can be embedded in a Gauss-Markov tree, then
\begin{align}
\label{firstcond}
|\rho_{ik}| & \ge |\rho_{ij} \rho_{jk}| \\
\intertext{and}
\label{secondcond}
\rho_{ik} \rho_{ij} \rho_{jk} & \ge 0
\end{align}
for all distinct $i$, $j$, and $k$. Conversely, if $N = 3$ and conditions
(\ref{firstcond}) and (\ref{secondcond}) hold for all distinct $i$
$j$, and $k$,
then $x_1,\ldots,x_N$ can be embedded in a Gauss-Markov tree.
\end{prop}
\prf
See Appendix~\ref{embedproof}.
\hfill $\Box$

\section{A Partial Converse}
\label{sec:discussion}

We have shown that if the source can be embedded in a Gauss-Markov
tree, then the separation-based scheme achieves the entire rate-distortion
region for the many-help-one problem. This raises the question of
whether the tree-embeddability
condition can be relaxed, or whether it is necessary in order for
the separation-based scheme to achieve the entire rate-distortion region.
We next show that it is reasonable to conjecture that tree-embeddability,
or a similar condition,  is a necessary and sufficient condition for
separation to achieve the entire rate-distortion region. Our argument
consists of two parts. 
\begin{itemize}
\item First, we provide an example that shows that 
separation does not always achieve the entire rate-distortion region 
for the many-help-one problem, which 
establishes that some added condition is required. 
\item We then establish a connection between this counterexample
  and the tree embeddability condition.
\end{itemize}

\subsection{Suboptimality of Separation}

We begin by showing that the separation-based scheme does not always
achieve the entire rate-distortion region for the many-help-one problem.
Consider the special case
of three sources ($N = 3$), where $x_1$ and $x_2$ have covariance
matrix
$$
\left[
\begin{array}{cc}
\sigma^2 & \rho \sigma^2 \\
\rho \sigma^2 & \sigma^2
\end{array}
\right] \quad 0 < \rho < 1.
$$
and where $x_3 = x_1 - x_2$. We shall assume that the goal is
to reproduce $x_3$ at the decoder and that $R_3 = 0$, i.e.,
the helpers completely shoulder the communication burden.

We shall focus in particular on the
asymptotic regime in which $\sigma^2$ is large and $\rho$ is near
one. Specifically, let
$$
\rho = 1 - \frac{1}{2 \sigma^2}
$$
and consider the behavior of the rate-distortion region as
$\sigma^2$ tends to infinity. Note that the variance of $x_3$
does not tend to infinity, and in fact equals one for any
positive value of $\sigma^2$, due to our choice of $\rho$.
In this regime, the separation-based scheme performs quite
poorly.

\begin{prop}
\label{BTisbad}
Let $0 < d < 1$ and  let
$R(\sigma^2,d)$ denote the minimum value of $R_1 + R_2$ such
that $(R_1,R_2,0,d)$ is in the rate-distortion region for the
separation-based scheme.  Then
$$
\lim_{\sigma^2 \rightarrow \infty} R(\sigma^2,d) = \infty.
$$
\end{prop}
\prf
Please see Appendix~\ref{BTbadproof}.
\hfill $\Box$

We now exhibit a scheme whose sum rate is bounded as
$\sigma^2$ tends to infinity. This scheme is simple in
the sense that it operates on individual samples, not
long blocks. Consider two lattices in $\real$,
\begin{align*}
\Lambda_i & = \{k \cdot 2^{-n} : k \in \ints \} \\
\Lambda_o & = \{k \cdot 2^{m} : k \in \ints \}.
\end{align*}
Let $Q_i(x)$ denote the lattice point in $\Lambda_i$ that
is closest to $x$; ties are broken arbitrarily. Let
$$
x \mod \Lambda_i = x - Q_i(x).
$$
Analogous definitions for $\Lambda_o$ are also in effect.

Let
$$
\tilde{x}_1(\ell) = Q_i(x_1(\ell)).
$$
For each time $\ell$, the first encoder communicates
$$
u_1(\ell) = \tilde{x}_1(\ell) \mod \Lambda_o
$$
to the decoder. This requires sending $n + m$ bits
per sample.
The second decoder operates analogously, yielding
a sum rate of $2(n+m)$ bits per sample.

The decoder uses
$$
\hat{x}_3(\ell) = \left[u_1(\ell) - u_2(\ell)\right] \mod \Lambda_o
$$
as its estimate for $x_3(\ell)$. 

\begin{prop}
\label{latticegood}
For any $d > 0$, if $m$ and $n$ are sufficiently large, then
$$
\E[(x_3(\ell) - \hat{x}_3(\ell))^2] \le d
$$

all $\ell$ and  all $\sigma^2$.
\end{prop}
\prf
Please see Appendix~\ref{latticeproof}.
\hfill $\Box$

Since $n$ and $m$ need not tend to infinity as $\sigma^2$ grows,
this simple scheme
beats the separation-based approach by an arbitrarily large amount
as $\sigma^2$ tends to infinity. The scheme can be
improved by using higher-dimensional lattices for $\Lambda_i$
and $\Lambda_o$. This has been explored by Krithivasan and Pradhan~\cite{KP07}.

Conceptually, the difference between the two schemes can
be understood as follows. Consider the binary expansion of
$x_1$. The quantity
$$
Q_i(x_1) \mod \Lambda_o
$$
can be computed from the sign of $x_1$ and the $m$ bits
to the left of the binary point and the $n + 1$ bits to the
right of the binary point. Thus, Proposition~\ref{latticegood}
shows that only these $n + m + 2$ bits are necessary for the purpose of
reproducing the difference $x_1 - x_2$. In particular, it
is not necesssary to send the bits that are more significant
than the block of $m$ to the left of the binary point.
As a result of using
a standard vector quantizer, however, the separation-based scheme
effectively sends these most significant bits. If the variances of $x_1$
and $x_2$
are large, this is inefficient.

\subsection{On the Necessity of the Tree Condition}

The previous section shows that the separation-based architecture
does not achieve the complete rate-distortion region when $x_1$
and $x_2$ are positively correlated and $x_3 = x_1 - x_2$,
at least when the variances of $x_1$ and $x_2$ are large and
their correlation coefficient is near one. This is also true
of the problem in which $x_1$ and $x_2$ are negatively correlated
and $x_3 = x_1 + x_2$. The defining feature of these two
examples is that if $\E[x_3|x_1,x_2] = a_1 x_1 + a_2 x_2$, then
\begin{equation}
\label{latticecond}
a_1 \cdot a_2 \cdot \E[x_1 x_2] < 0.
\end{equation}
We next show that for $N = 3$, if the sources cannot be embedded
in a Gauss-Markov tree, then this condition holds, except
for a possible relabeling.

\begin{prop}
\label{prop:converse}
For $N = 3$, if $x_1$, $x_2$, and $x_3$ cannot be embedded in
a Gauss-Markov tree, then~(\ref{latticecond}) holds for some
relabeling of $x_1$, $x_2$, and $x_3$.
\end{prop}

\prf
Please see Appendix~\ref{app:converse}.
\hfill $\Box$

\appendix

\section{Proof of Lemma~\ref{lemma:outer}}
\label{app:outer}

\begin{figure}[ht]
\begin{center}
\scalebox{1.0}{\input{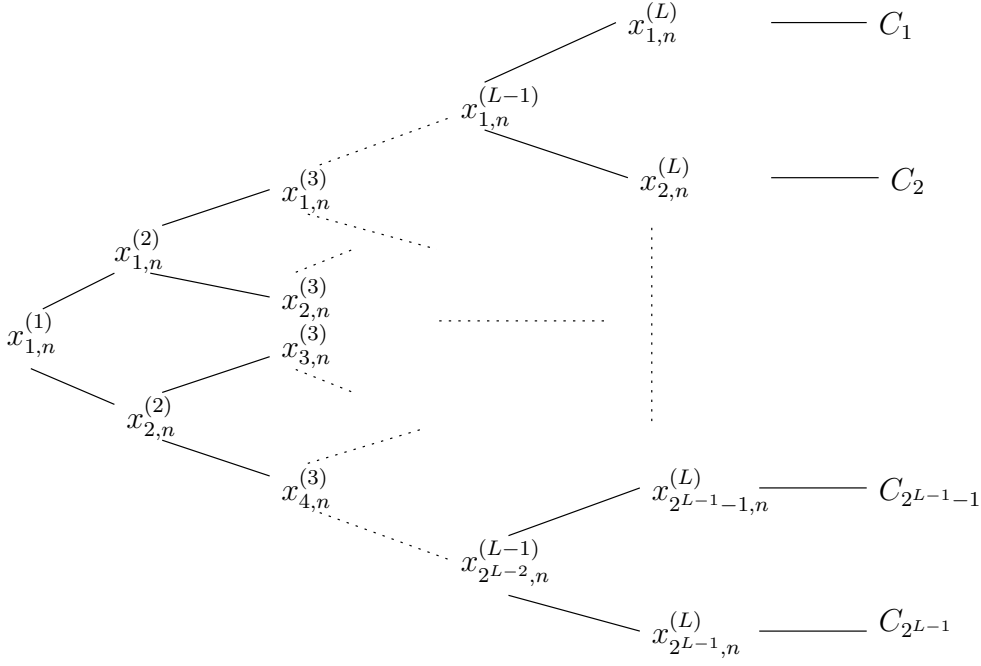}}
\end{center}
\caption{The tree structure with the encoder outputs over a block of length $n$.}
 \label{fig:tree+distributed}
\end{figure}

Consider any encoding-decoding procedure that achieves the rate-distortion
tuple $$(R_1,R_2,\ldots,R_{2^{L-1}},d)$$
for
the binary tree structure problem over a block of time of length $n$.
Let the discrete set $C_i$ denote the output of encoder $i$
(for $i=1,\ldots ,2^{L-1}$). We have that
\beqa
R_i & \ge & \frac{1}{n} \log |C_i|,\quad i=1,\ldots ,2^{L-1} \\
d & \ge & \frac{1}{n} \sum_{m=1}^n {\rm Var}\lp x_1^{(1)}(m) | \ve{C}\rp. 
\label{eq:distortionbound}
\eeqa
Here we have denoted
\beq
\ve{C} \df \lbr C_1,\ldots ,C_{2^{L-1}}\rbr,
\eeq
the set of all the encoder outputs.
Further, the distributed nature of encoding imposes natural Markov chain conditions
on the encoder outputs with respect to the observations. These Markov chain
conditions are described in Figure~\ref{fig:tree+distributed}.

Recall our earlier definition of the {\em ancestors} set $\C{A}^{(k)}$ (c.f.\
Equation~\eqref{eq:ancestors})
\beq
\label{eq:ancestors_again}
\C{A}^{(k)}  \df \left\{ i : \C{O}(x_i^{(k)}) \cap \C{A} \not= \Phi  \right\},
\eeq
where $\Phi$ is the null set. Now define
\beq
\ve{x}_{\C{A},n}^{(k)}  \df  \left\{ x_{i,n}^{(k)}: i \in \C{A}^{(k)}\rbr.
\eeq
Our outer bound will consider arbitrary subsets $\C{A}$ of $\lbr 1,\ldots ,2^{L-1}\rbr$.
Denote the set
\beq
\ve{C}_{\C{A}} \df \lbr C_i: i\in\C{A}\rbr.
\eeq
The sum of any subset $\C{A}$ of the encoder rates satisfies
\beqa
n\sum_{i\in\C{A}} R_i & = & \sum_{i\in\C{A}} \log |C_i| \nonumber \\
\nonumber &\geq & \sum_{i \in \C{A}} H(C_i) \\
\nonumber &\geq & H\lp\ve{C}_{\C{A}}\rp \\
\nonumber & \geq &H\lp\ve{C}_{\C{A}}|\ve{C}_{\C{A}^c}\rp\\
\nonumber& = &I\lp\ve{x}_{\C{A},n}^{(L)};\ve{C}_{\C{A}}|\ve{C}_{\C{A}^c}\rp\\
\label{eq:eqa} &\stackrel{(a)}{=} & I\lp\ve{x}_{\C{A},n}^{(1)},\cdots\ve{x}_{\C{A},n}^{(L-1)},\ve{x}_{\C{A},n}^{(L)};
\ve{C}_{\C{A}}|\ve{C}_{\C{A}^c}\rp\\
\label{eq:eqb} &\stackrel{(b)}{=} &\sum_{k=1}^{L} I\lp\ve{x}_{\C{A},n}^{(k)};\ve{C}_{\C{A}} | \ve{x}_{\C{A},n}^{(k-1)},\ve{C}_{\C{A}^c}\rp\\
 &\stackrel{(c)}{=} & \sum_{k=1}^{L}\left(I\lp\ve{x}_{\C{A},n}^{(k)};\ve{C}| \ve{x}_{\C{A},n}^{(k-1)}\rp -
 I\lp\ve{x}_{\C{A},n}^{(k)};\ve{C}_{\C{A}^c} | \ve{x}_{\C{A},n}^{(k-1)}\rp \right).\label{eq:eqc}
\eeqa
Here each of the steps $(a)$, $(b)$, and $(c)$ follow from
 the Markov chain conditions described in Figure~\ref{fig:tree+distributed}.
We use the chain rule to expand each of the mutual information terms in the
lower bound of Equation~\eqref{eq:eqc}:
\beqa
I\lp\ve{x}_{\C{A},n}^{(k)};\ve{C}| \ve{x}_{\C{A},n}^{(k-1)}\rp & = & \sum_{i\in\C{A}^{(k)}}
I\lp x_{i,n}^{(k)}; \ve{C} | \ve{x}_{\C{A},n}^{(k-1)}, x_{j,n}^{(k)}, j < i, j\in\C{A}^{(k)}\rp \\
& = & \sum_{i\in\C{A}^{(k)}}I\lp  x_{i,n}^{(k)}; \ve{C} | x^{(k-1)}_{\lfloor\frac{i+1}{2}\rfloor,n}\rp,\label{eq:chnrule1}
\eeqa
and
\beqa
 I\lp\ve{x}_{\C{A},n}^{(k)};\ve{C}_{\C{A}^c} | \ve{x}_{\C{A},n}^{(k-1)}\rp
 & = &\sum_{i\in\C{A}^{(k)}} I\lp  x_{i,n}^{(k)}; \ve{C}_{\C{A}^c} | \ve{x}_{\C{A},n}^{(k-1)}, x_{j,n}^{(k)},
       j < i, j\in\C{A}^{(k)}\rp \\
 & = & \sum_{i\in\C{A}^{(k)}}I\lp  x_{i,n}^{(k)}; \ve{C}_{\C{A}^c} | x^{(k-1)}_{\lfloor\frac{i+1}{2}\rfloor,n}\rp\label{eq:chnrule2}
 \eeqa
Here both Equations~\eqref{eq:chnrule1} and~\eqref{eq:chnrule2} follow from the Markov chain
conditions described in  Figure~\ref{fig:tree+distributed}.
Denote by
\beq\label{eq:rikdefn}
r_i^{(k)} \df \frac{1}{n} I\lp  x_{i,n}^{(k)}; \ve{C} | x^{(k-1)}_{\lfloor\frac{i+1}{2}\rfloor,n}\rp,
\eeq
the term inside the summation in Equation~\eqref{eq:chnrule1}. 
Then $r_1^{(1)}$ is the number of bits per sample that the encoders
send about the root of the tree and $r_i^{(k)}$ for $k > 1$ can be
interpreted as the number of bits per sample that the encoders 
use to represent the noise
introduced at node $x_i^{(k)}$.
 We will upper bound the terms
inside the summation in Equation~\eqref{eq:chnrule2} 
in terms of these quantities. To do
this, we start with  a central preliminary lemma.
\subsection{A Preliminary Lemma}
\begin{figure}[ht]
\begin{center}
\scalebox{1.2}{\input{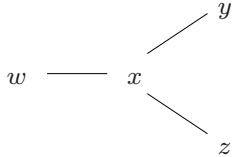}}
\end{center}
\caption{The Markov chain conditions.} \label{fig:wxyz_again}
\end{figure}

Consider four memoryless jointly Gaussian
random processes $w(m), x(m), y(m), z(m),
m=1,\ldots,n$. They are identically jointly distributed in the (time) index $m$.
At any given time index $m$, their joint distribution satisfies the Markov chain
conditions implied in Figure~\ref{fig:wxyz_again}.
Then we can write, for all $m=1,\ldots ,n$,
\begin{eqnarray*}
x(m) & = & \alpha_{xw}w(m) + n_0(m),\\
y(m) & = & \alpha_{yx}x(m) + n_1(m),\\
z(m) & = & \alpha_{zx}x(m) + n_2(m),
\end{eqnarray*}
for some real $\alpha_{xw},\alpha_{yx},\alpha_{zx}$.
Here $n_0(m),~n_1(m),~n_2(m), m=1\ldots ,n$, are i.i.d.\ in time
 and independent of each other and independent of the process $w(m), m=1,
 \ldots ,n$. Further, the random variables $n_0(m),n_1(m),n_2(m),w(m)$
 at any time index $n$ are  $\C{N}(0,\sigma_{n_0}^2), \C{N}(0,\sigma_{n_1}^2),
\C{N}(0,\sigma_{n_2}^2)$, and  $\C{N}(0,\sigma_{w}^2)$ respectively.

Write the vectors
\beqa
w_n & = & \Lbr w(1), \ldots ,w(n)\Rbr \\
x_n & = & \Lbr x(1), \ldots ,x(n)\Rbr \\
y_n & = & \Lbr y(1), \ldots ,y(n)\Rbr \\
z_n & = & \Lbr z(1), \ldots ,z(n)\Rbr.
\eeqa
Consider two random variables $C_1,C_2$ that satisfy the following
two Markov chain conditions:
\beqa
\lp w_n, x_n, z_n, C_2\rp\quad \markov & y_n & \markov \quad C_1, \label{eq:m1}\\
\lp w_n, x_n, y_n, C_1\rp\quad \markov & z_n & \markov \quad C_2,\label{eq:m2}
\eeqa
Our first inequality concerns this Markov chain condition.
We intentionally use notation similar to that introduced in 
Section~\ref{sec:tree_conv}.
\begin{lemma}\label{lemma:3.2}
Define
\begin{eqnarray*}
r_1 & \df & \frac{1}{n}I(y_n;C_1|x_n),\\
r_2 & \df & \frac{1}{n}I(z_n;C_2|x_n),\\
f_{x}\lp r_1,r_2\rp &\df& \frac{1}{2}\log\lp 1 + \frac{\alpha_{yx}^2 \sigma^2_{n_0}}{\sigma^2_{n_1}}
\lp 1 - e^{-2r_1}\rp + \frac{\alpha^2_{zx} \sigma^2_{n_0}}{\sigma^2_{n_2}}
\lp 1 - e^{-2r_2}\rp \rp. \label{eq:fx}
\end{eqnarray*}
Then
\begin{eqnarray}\label{eq:ineq1}
\frac{1}{n}I(x_n;C_1,C_2|w_n) & \leq &
f_x(r_1,r_2),\\\label{eq:ineq2} \frac{1}{n}I(x_n;C_1|w_n) & \leq &
f_x(r_1,0),\\\label{eq:ineq3} \frac{1}{n}I(x_n;C_2|w_n) & \leq &
f_x(0,r_2).
\end{eqnarray}
\end{lemma}
{\bf Proof}:
This lemma is a conditional version (conditioned on $w_n$) of Lemma
3 in \cite{Oohama05}. The proof follows ``mutatis mutandis" that of 
Lemma 3 in
\cite{Oohama05}; the only extra fact needed is that conditioned 
on {\em any} realization of
$w_n$, $(x_n,y_n,z_n)$ are jointly Gaussian with their original
variances and 
$(x_n, y_n, z_n, C_1, C_2)$ satisfies the Markov condition
$$
C_1 \markov y_n \markov x_n \markov z_n \markov C_2.
$$
Specifically, suppose first that
$\alpha_{yx}$ and $\alpha_{zx}$ are nonzero.
For any realization of $w_n$, say $\tilde{w}_n$,
Oohama~\cite[Lemma~3]{Oohama05} has shown that
$$
\frac{1}{n} I(x_n;C_1,C_2|w_n = \tilde{w}_n)
  \le f_x(r_1,r_2).
$$
By averaging the left-hand side over $\tilde{w}_n$, we obtain
(\ref{eq:ineq1}). The proofs of~(\ref{eq:ineq2}) and~(\ref{eq:ineq3})
are similar. If both $\alpha_{yx}$ and $\alpha_{zx}$ are
zero, then the result is trivial. If, say, only $\alpha_{yx}$
is zero, then $I(x_n;C_1,C_2|w_n) = I(x_n;C_2|w_n)$ and
(\ref{eq:ineq1}) follows from~(\ref{eq:ineq3}).
\hfill $\Box$

\subsubsection{Sufficient Conditions for Equality}
\label{app:equality}
It is useful to observe the conditions for equality 
in~\eqref{eq:ineq1},~\eqref{eq:ineq2} and~\eqref{eq:ineq3}: suppose
\beq\label{eq:achoice}
C_k = \Lbr u_k(1),\ldots ,u_k(n)\Rbr,\quad k=1,2.
\eeq
Here
\begin{eqnarray*}
u_1(m) = \alpha_1 y(m) + v_1(m),\quad m=1,\cdots,n,\\
u_2(m) = \alpha_2 z(m) + v_2(m),\quad m=1,\cdots,n,
\end{eqnarray*}
where $v_1(m)$ and $v_2(m)$ are Gaussian and independent of each other and of
$w_n,x_n,y_n,z_n$ and are i.i.d.\ in the time index $m$.
Then it is verified directly that with this choice of $C_1,C_2$
(c.f.\ Equation~\eqref{eq:achoice}) the inequalities
 in Equations~\eqref{eq:ineq1},~\eqref{eq:ineq2} and~\eqref{eq:ineq3}
 are all simultaneously met with
equality (this verification is also done in \cite{Oohama05,PTR04}).
This fact will be used later to show that the achievable
region of the separation-based inner bound coincides with the outer bound.

\subsubsection{An Important Instance}\label{sec:important_instance}
Of specific interest to us will be  the following association of the random
variables in Figure~\ref{fig:wxyz_again} to the binary tree structure in
Figure~\ref{fig:tree}:
fix $1\leq k\leq L-1$ and $1\leq i\leq 2^{k-1}$. Then let
\beqa
x &=& x_i^{(k)}\\
y & = & x_{2i-1}^{(k+1)} \\
z & = & x_{2i}^{(k+1)}\\
w & = & x^{(k-1)}_{\lfloor \frac{i+1}{2}\rfloor}.
\eeqa
With this association, denote the function corresponding to $f_x$ in Equation~\eqref{eq:fx}
 by $f_{x_i^{(k)}}$:
\beq \label{eq:fxik_again} f_{x_i^{(k)}}\lp r_1,r_2\rp \df
\frac{1}{2}\log\lp 1 +
\frac{\alpha_{2i-1}^{(k+1)}
\sigma^2_{n_i^{(k)}}}{\sigma^2_{n_{2i-1}^{(k+1)}}} \lp 1 -
e^{-2r_1}\rp + \frac{\alpha_{2i}^{(k+1)}
   \sigma^2_{n_i^{(k)}}}{\sigma^2_{n_{2i}^{(k+1)}}} \lp
1 - e^{-2r_2}\rp \rp, \quad r_1,r_2\geq 0. \eeq Indeed, this is the
same notation as that introduced in Section~\ref{sec:tree_conv}
(c.f.\ Equation~\eqref{eq:fxik_defn}).

\subsection{An Iteration Lemma}
As an immediate application of the preliminary lemma derived in the previous section,
consider any subset $\C{A} \subseteq \lbr 1,\ldots ,2^{L-1}\rbr$. Fix $ 1\leq k\leq L-1$ and
$1\leq i\leq 2^{k-1}$. For simplicity of notation, let us suppose that
 $x_{1}^{(0)}$ is a zero random variable.
\begin{lemma}\label{lemma:3.2again}
\beq\label{eq:fxik}
\frac{1}{n}I\lp x_{i,n}^{(k)}; \ve{C}_{\C{A}}|x_{\lfloor \frac{i+1}{2}
\rfloor, n}^{(k-1)}\rp \leq
f_{x_{i}^{(k)}}\lp
\frac{1}{n}I\lp x_{2i-1,n}^{(k+1)}; \ve{C}_{\C{A}}|x_{i,n}^{(k)}\rp,
\frac{1}{n}I\lp x_{2i,n}^{(k+1)}; \ve{C}_{\C{A}}|x_{i,n}^{(k)}\rp 
\rp.
\eeq
\end{lemma}
{\bf Proof}:
For any node $x_i^{(k)}$, recall  the set of associated
{\em observations} defined as (c.f.\ Equation~\eqref{eq:observe})
\begin{eqnarray*}
\C{O}\lp x_i^{(k)}\rp & = & \left\{j: \frac{2^{L}(i-1)}{2^{k}} < j \leq
\frac{2^{L}i}{2^{k}} \right\}.\label{eq:observe_again}
\end{eqnarray*}
With this definition, we observe that
\begin{eqnarray*}
\C{O}\lp x_i^{(k)}\rp&=& \C{O}\lp x_{2i-1}^{(k+1)}\rp\cup
\C{O}\lp x_{2i}^{(k+1)}\rp, \\ I \left(x_{i,n}^{(k)};
\ve{C}_{\C{A}}|x_{\lfloor \frac{i+1}{2} \rfloor, n}^{(k-1)}\right)&=&
I \left(x_{i,n}^{(k)}; \ve{C}_{\C{A}\cap\C{O}\lp x_i^{(k)}\rp}|x_{\lfloor
\frac{i+1}{2} \rfloor, n}^{(k-1)}\right).
\end{eqnarray*}
Then we only need to invoke 
Lemma \ref{lemma:3.2} with the
following random variables:
\begin{eqnarray*}
w_n & = & x_{\lfloor \frac{i+1}{2} \rfloor, n}^{(k-1)},\\
x_n & = & x_{i,n}^{(k)},\\
y_n & = & x_{2i-1,n}^{(k+1)},\\
z_n & = & x_{2i,n}^{(k+1)}, \\
C_1 & = & \ve{C}_{\C{A}\cap \C{O}\lp x_{2i-1}^{(k+1)}\rp},\\
C_2 & = & \ve{C}_{\C{A}\cap \C{O}\lp x_{2i}^{(k+1)}\rp}.
\end{eqnarray*}
This completes the proof.\hfill $\Box$

Observe that the parameters inside the function $f_{x_i^{(k)}}\lp \cdot, \cdot\rp$ are
themselves of the type of the term in the left hand side of Equation~\eqref{eq:fxik}.
Then, we can repeatedly apply Lemma~\ref{lemma:3.2again}.
As an example, we have for $k \leq L-2$ and $1\leq i\leq 2^{k-1}$, the two parameters
of $f_{x_i^{(k)}}$ in Equation~\eqref{eq:fxik} are upper bounded by
\beqa
\frac{1}{n} I\lp x_{2i-1,n}^{(k+1)} ; 
   \ve{C}_{\C{A}} | x_{i,n}^{(k)}\rp  &\leq&
f_{x_{2i-1}^{(k+1)}}\lp 
\frac{1}{n} I\lp x_{4i-3,n}^{(k+2)}; \ve{C}_{\C{A}} | x_{2i-1,n}^{(k+1)}\rp,
\frac{1}{n} I\lp x_{4i-2,n}^{(k+2)}; \ve{C}_{\C{A}} | x_{2i-1,n}^{(k+1)}\rp
    \rp. \label{eq:useagain2}
 \\
\frac{1}{n} I\lp x_{2i,n}^{(k+1)} ; \ve{C}_{\C{A}} | x_{i,n}^{(k)}\rp &\leq&
f_{x_{2i}^{(k+1)}}\lp 
\frac{1}{n} I\lp x_{4i-1,n}^{(k+2)}; \ve{C}_{\C{A}} | x_{2i,n}^{(k+1)}\rp,
\frac{1}{n} I\lp x_{4i,n}^{(k+2)}; \ve{C}_{\C{A}} | x_{2i,n}^{(k+1)}\rp
\rp. \label{eq:useagain1}
\eeqa
Now the function $f_{x_i^{(k)}}\lp \cdot, \cdot\rp$ is monotonically increasing
in both of
its parameters (this is true for each $1\leq k\leq L-1$ and $1\leq i\leq 2^{k-1}$).
So, we can combine Equations~\eqref{eq:fxik},
\eqref{eq:useagain1}
and~\eqref{eq:useagain2} to get
\beqa
\frac{1}{n} I(x_{i,n}^{(k)}; \ve{C}_{\C{A}}|x_{\lfloor \frac{i+1}{2}
\rfloor, n}^{(k-1)}) & \leq &  f_{x_{i}^{(k)}}\lp
f_{x_{2i-1}^{(k+1)}}\lp 
\frac{1}{n} I\lp x_{4i-3,n}^{(k+2)}; \ve{C}_{\C{A}} | x_{2i-1,n}^{(k+1)}\rp,
\frac{1}{n} I\lp x_{4i-2,n}^{(k+2)}; \ve{C}_{\C{A}} | x_{2i-1,n}^{(k+1)}\rp
   \rp,\right. \nonumber \\
  &&\quad \left. 
    f_{x_{2i}^{(k+1)}}\lp 
  \frac{1}{n} I\lp x_{4i-1,n}^{(k+2)}; \ve{C}_{\C{A}} | x_{2i,n}^{(k+1)}\rp,
  \frac{1}{n} I\lp x_{4i,n}^{(k+2)}; \ve{C}_{\C{A}} | x_{2i,n}^{(k+1)}\rp
   \rp\rp.
\eeqa
The stage is now set to recursively apply Lemma~\ref{lemma:3.2again}. Continuing this process
until the boundary conditions are met, we arrive at
\beq\label{eq:fAx}
\frac{1}{n} I\lp x_{i,n}^{(k)} ; \ve{C}_{\C{A}} | x_{\lfloor\frac{i+1}{2}\rfloor ,n}^{(k-1)}\rp
\leq f_{x_i^{(k)}}^{\C{A}}\lp \ve{r}_{\C{A}}\lp x_{i}^{(k)}\rp\rp.
\eeq
Here the set $\ve{r}_{\C{A}}\lp x_i^{(k)}\rp$ is defined as in Equation~\eqref{eq:raxik}:
\begin{eqnarray}\nonumber
\ve{r}_{\C{A}}\lp x_i^{(k)}\rp & = & \left\{r_j^{(l)}:~x_j^{(l)} \in
\C{T}\lp x_i^{(k)}\rp,~ \C{O}(x_j^{(l)}) \subset \C{A}, \right.
\\ \nonumber &&  \not\exists~ x_a^{(b)}~\in \C{T}\lp x_i^{(k)}\rp
~\mbox{with}~\C{O}\lp x_a^{(b)}\rp \subset \C{A},\\\label{eq:raxik_again}
&&\left.\mbox{and}~x_j^{(l)}~\in~\C{R}\lp x_a^{(b)}\rp\cup\C{L}(x_a^{(b)})
\right\}.
\end{eqnarray}
The function $f_{x_i^{(k)}}^{\C{A}}(\cdot)$ was also defined  in
Section~\ref{sec:tree_conv}.
\subsection{Putting Them Together}
We are now ready to complete the proof of Lemma~\ref{lemma:outer}.
First, we substitute Equation~\eqref{eq:fAx} in Equation~\eqref{eq:chnrule2} to get
\beq\label{eq:chnrule2_revise}
I\lp\ve{x}_{\C{A},n}^{(k)};\ve{C}_{\C{A}^c} | \ve{x}_{\C{A},n}^{(k-1)}\rp \leq
  \sum_{i\in\C{A}^{(k)}} f_{x_i^{(k)}}^{\C{A}^c}
   \lp \ve{r}_{\C{A}^c}\lp x_{i}^{(k)}\rp\rp.
\eeq
Combining Equation~\eqref{eq:chnrule2_revise} with Equations~\eqref{eq:chnrule1}
and~\eqref{eq:rikdefn}, we can rewrite the inequality in Equation~\eqref{eq:eqc} as
\beq\label{eq:finalbound}
\sum_{i\in\C{A}} R_i  \geq  \sum_{k=1}^{L}\sum_{i
\in \C{A}^{(k)}} \left(r_i^{(k)}
-f_{x_{i}^{(k)}}^{\C{A}^c}\left(\ve{r}_{\C{A}^c}(x_i^{(k)})\right)\right).
\eeq
The quantities $r_i^{(k)}$ satisfy other natural inequalities as well:
\begin{itemize}
\item Supposing that $\C{A}$ equals 
the entire set $\lbr 1,2,\ldots ,2^{L-1}\rbr$ and
substituting in Lemma~\ref{lemma:3.2again} we have
\beq\label{eq:cnst1}
r_i^{(k)} \leq f_{x_i}^{(k+1)}\lp r_{2i-1}^{(k+1)}, r_{2i}^{(k+1)}\rp.
\eeq
\item By direct calculation we also have
\beqa
r_{1}^{(1)} & = & \frac{1}{n}I\lp x_{1,n}^{(1)};\ve{C}\rp\\
& = & \frac{1}{n} h\lp x_{1,n}^{(1)}\rp - \frac{1}{n}
   h\lp x_{1,n}^{(1)} | \ve{C}\rp \\
& \geq & \frac{1}{2}\log\lp 2 \pi e \sigma^2_{x_1^{(1)}}\rp -
   \frac{1}{n}
   h\lp x_{1,n}^{(1)} - \E\Lbr x_{1,n}^{(1)}|\ve{C}
   \Rbr \rp \label{eq:mmseb} \\
& \geq & \frac{1}{2}\log\lp \sigma^2_{x_1^{(1)}}\rp - \frac{1}{2n}\log\det\lp
{\rm Covar}\lp x_{1,n}^{(1)} - \E\Lbr x_{1,n}^{(1)}| \ve{C}\Rbr\rp\rp \label{eq:covari}\\
& \geq & \frac{1}{2}\log\lp \sigma^2_{x_1^{(1)}}\rp - \frac{1}{2}\log\lp \frac{1}{n}{\rm Trace}\lp
        {\rm Covar}\lp x_{1,n}^{(1)} - \E\Lbr x_{1,n}^{(1)}| \ve{C}\Rbr\rp\rp\rp \label{eq:Hadamard} \\
& = & \frac{1}{2}\log\lp \sigma^2_{x_1^{(1)}}\rp - \frac{1}{2}\log\lp \frac{1}{n}\sum_{m=1}^n {\rm Var}\lp
x_{1}^{(1)}(m)|\ve{C}\rp \rp \\
& \geq & \frac{1}{2}\log\frac{\sigma_{x_1^{(1)}}^2}{d}. \label{eq:distb}
\eeqa
where:
\begin{itemize}
\item Equation~\eqref{eq:mmseb} follows from the fact that conditioning only reduces the differential
entropy;
\item Equation~\eqref{eq:covari} is the usual bound on the differential entropy of a vector by
the determinant of its covariance matrix;
\item Equation~\eqref{eq:Hadamard} follows from the Hadamard inequality on the determinant of
a positive definite matrix in terms of its trace;
\item Equation~\eqref{eq:distb} follows from the fact that the encoder outputs describe the
original root node of the tree with sufficiently small quadratic fidelity
(c.f.\ Equation~\eqref{eq:distortionbound}).
\end{itemize}
\end{itemize}
Based on Equations~\eqref{eq:cnst1} and~\eqref{eq:distb} we see that the set of $r_i^{(k)}$
indeed belong to the set $\C{F}_r(d)$ defined in Equation~\eqref{eq:frd}. Combining this fact
with the key inequality in Equation~\eqref{eq:finalbound}, we have completed the proof of the
outer bound in Lemma~\ref{lemma:outer}. \hfill $\Box$
\section{Proof of Lemma~\ref{lemma:matchup}}
\label{app:matchup}
Since we know that
\beq
\Co(\C{RD}_{\rm{in}}) \subset \C{RD}_{\rm{out}},
\eeq
it suffices to prove that for any $d$ and any
componentwise nonnegative vector $(\alpha_1,\ldots,
\alpha_{2^{L-1}})$,
\begin{eqnarray*}
\inf_{\ve{R}: (\ve{R},d) \in \C{RD}_{\rm{out}}} \sum_{i=1}^{2^{L-1}}
\alpha_i R_i & \ge & \inf_{\ve{R}: (\ve{R},d) \in \C{RD}_{\rm{in}}}
\sum_{i=1}^{2^{L-1}} \alpha_i R_i.
\end{eqnarray*}
We will assume that $\alpha_1 \leq \alpha_2 \leq \cdots \leq
\alpha_{2^{L-1}}$. The proof for the other orderings is similar.
We will also use the convention $\alpha_0 = 0$. Now for
any $R_1,\ldots,R_{2^{L-1}}$,
\begin{eqnarray*}
\sum_{i=1}^{2^{L-1}} \alpha_i R_i & = & \alpha_1
\sum_{i=1}^{2^{L-1}} R_i + (\alpha_2-\alpha_1) \sum_{i=2}^{2^{L-1}}
R_i + \\ && \cdots + (\alpha_{2^{L-1}}-\alpha_{2^{L-1}-1})
R_{2^{L-1}}, \\
& = &
\sum_{j=1}^{2^{L-1}}(\alpha_j-\alpha_{j-1})\sum_{i=j}^{2^{L-1}} R_i
\end{eqnarray*}
Thus
$$
\inf_{\ve{R}: (\ve{R},d) \in \C{RD}_\mathrm{out}}
  \sum_{i = 1}^{2^{L-1}} \alpha_i R_i =
\inf_{\ve{R}: (\ve{R},d) \in \C{RD}_\mathrm{out}}
  \sum_{j = 1}^{2^{L-1}} (\alpha_j - \alpha_{j-1})
   \sum_{i = 1}^{2^{L-1}} R_i.
$$
Let $\epsilon > 0$. Then there exists $s \in \C{F}_r(d)$ and $\ve{R}^*$
such that
$$
  \sum_{j = 1}^{2^{L-1}} (\alpha_j - \alpha_{j-1})
   \sum_{i = 1}^{2^{L-1}} R_i^* \le
  \inf_{\ve{R}: (\ve{R},d) \in \C{RD}_\mathrm{out}}
  \sum_{j = 1}^{2^{L-1}} (\alpha_j - \alpha_{j-1})
   \sum_{i = 1}^{2^{L-1}} R_i + \epsilon
$$
and
$$
\sum_{i \in \C{A}} R_i^* \ge \sum_{k = 1}^L \sum_{i \in \C{A}^{(k)}}
  \lp s_i^{(k)} - f_{x_i^{(k)}}^{\C{A}^c}(\ve{s}_{\C{A}^c}(x_i^{(k)}))\rp
$$
for all $A$. Let
$$
\C{A}_j = \{j, \ldots, 2^{L-1}\} \cap \{i: s_i^{(L)} > 0\}.
$$
Then
\begin{align*}
  \sum_{j = 1}^{2^{L-1}} (\alpha_j - \alpha_{j-1})
   \sum_{i = 1}^{2^{L-1}} R_i^* 
   & \ge \sum_{j = 1}^{2^{L-1}} (\alpha_j - \alpha_{j-1}) 
          \sum_{i \in \C{A}_j} R_i^* \\
   & \ge \sum_{j = 1}^{2^{L-1}} (\alpha_j - \alpha_{j-1}) 
        \sum_{k = 1}^L  \sum_{i \in \C{A}_j^{(k)}} 
     (s_i^{(k)} - f_{x_i^{(k)}}^{\C{A}_j^c}(\ve{s}_{\C{A}^c}(x_i^{(k)}))) \\
   & \ge \inf \sum_{j = 1}^{2^{L-1}} (\alpha_j - \alpha_{j-1}) 
        \sum_{k = 1}^L  \sum_{i \in \C{A}_j^{(k)}} 
     (r_i^{(k)} - f_{x_i^{(k)}}^{\C{A}_j^c}(\ve{r}_{\C{A}^c}(x_i^{(k)}))),
\end{align*}
where the infimum is over all $\ve{r}$ in $\C{F}_r(d)$ such that
$r_i^{(L)} = 0$ if and only if $s_i^{(L)} = 0$. Then there exists
$\tilde{\ve{s}} \in \C{F}_r(d)$ such that $\tilde{\ve{s}}^{(L)}_i = 0$
if and only if $\ve{s}_i^{(L)} = 0$ and
\begin{multline}
\label{eq:lowerboundmin}
   \sum_{j = 1}^{2^{L-1}} (\alpha_j - \alpha_{j-1}) 
        \sum_{k = 1}^L  \sum_{i \in \C{A}_j^{(k)}} 
     (\tilde{s}_i^{(k)} - 
   f_{x_i^{(k)}}^{\C{A}_j^c}(\tilde{\ve{s}}_{\C{A}^c}(x_i^{(k)}))) \\
  \le \inf \sum_{j = 1}^{2^{L-1}} (\alpha_j - \alpha_{j-1}) 
        \sum_{k = 1}^L  \sum_{i \in \C{A}_j^{(k)}} 
     (r_i^{(k)} - f_{x_i^{(k)}}^{\C{A}_j^c}(\ve{r}_{\C{A}_j^c}(x_i^{(k)})))
   + \epsilon
\end{multline}
and the $\tilde{\ve{s}}$ minimize
\begin{equation}
\label{eq:sumb}
\sum_{k = 1}^L \sum_{i = 1}^{2^{k-1}} \tilde{s}_i^{(k)}.
\end{equation}
Now since the $\tilde{s}_i^{(k)}$ are in $\C{F}_r(d)$, we have
\begin{align}
\label{btight1}
\tilde{s}_1^{(1)} & \ge \frac{1}{2}\log\frac{\sigma_{x_1^{(1)}}^2}{d} \\
\label{btight2}
\tilde{s}_{i}^{(k)} & 
   \leq f_{x_{i}^{(k)}}(s_{2i - 1}^{(k+1)}, s_{2i}^{(k+1)}).
\end{align}
We will show that both of these inequalities must actually be
equalities. Since the left-hand side of (\ref{eq:lowerboundmin}) is
monotonically decreasing in $s_1^{(1)}$ and  the $s_i^{(k)}$ 
minimize (\ref{eq:sumb}), it follows that the $s_1^{(1)}$ inequality
must be tight.

Next suppose that 
\beq
\label{eq:contradict}
\tilde{s}_{m}^{(n)} < f_{x_{m}^{(n)}}(\tilde{s}_{2m-1}^{(n+1)}, 
   \tilde{s}_{2m}^{(n+1)})
\eeq
for some non-leaf node $x_m^{(n)}$. We will show that this is
incompatible with the assumption that the $\tilde{s}_i^{(k)}$ minimize
(\ref{eq:sumb}).
Without loss of generality,
we may assume that none of the children of $x_m^{(n)}$ have a
strict inequality in (\ref{btight2}).
In order for (\ref{eq:contradict}) to hold, $\tilde{s}_j^{(L)}$
must be positive for at least one leaf variable $x_j^{(L)}$ under
$x_m^{(n)}$.
Consider the leaf variable $x_{\hat{m}}^{(L)}$
under $x_{m}^{(n)}$ with the largest index $\hat{m}$
such that $\tilde{s}_{\hat{m}}^{(L)}$
is positive:
$$
\hat{m} = \arg \max\left\{\frac{2^L (m - 1)}{2^n} < 
  j \le \frac{2^L m}{2^n} : \tilde{s}_j^{(L)} > 0\right\}.
$$
Then consider the descendant of $x_m^{(n)}$, $x_{\tilde{m}}^{(n+1)}$, that
leads to the leaf variable $x_{\hat{m}}^{(L)}$. Note that we must have
$\tilde{s}_{\tilde{m}}^{(n+1)} > 0$.

Suppose that we decrease $\tilde{s}_{\tilde{m}}^{(n+1)}$ by
a slight amount such that~(\ref{eq:contradict}) still holds.
Fix a $j$ in $\{1,\ldots,2^{L-1}\}$ and consider the sum
\begin{equation}
\label{eq:stildesum}
    \sum_{k=1}^{L}\sum_{i
        \in \C{A}_j^{(k)}} \left(\tilde{s}_i^{(k)}
    -f_{x_{i}^{(k)}}^{\C{A}_j^c}
     \lp\tilde{\ve{s}}_{\C{A}^c_j}(x_i^{(k)})\rp\rp,
\end{equation}
and recall that
$$
\C{A}_j = \{j,\ldots,2^{L-1}\} \cap \{i : \tilde{s}_i^{L} > 0\}.
$$
Now if $j > \hat{m}$, then all of the observations under $x_m^{(n)}$
are in $\C{A}_j^c$, which implies that the sum in (\ref{eq:stildesum})
does not depend on $\tilde{s}_{\tilde{m}}^{(n+1)}$. 
On the other hand, if $j \le \hat{m}$, then
not all of the observations under $\tilde{s}_{\tilde{m}}^{(n+1)}$ are in
$\C{A}_j^c$, and so
$$
\tilde{s}_{\tilde{m}}^{(n+1)} 
   \notin \tilde{\ve{s}}_{\C{A}_j^c}\lp x_i^{(k)} \rp
$$
for all $x_i^{(k)}$. It follows that the objective in (\ref{eq:lowerboundmin})
is not increased while the sum in (\ref{eq:sumb}) is reduced by
decreasing $\tilde{s}_{\tilde{m}}^{(n+1)}$,
which is a contradiction. Thus (\ref{eq:contradict})
cannot hold at any non-leaf nodes in the tree. We have thus shown
that equality
must hold in (\ref{btight1}) and (\ref{btight2}).

We are now in a position to show that
$$
   \sum_{j = 1}^{2^{L-1}} (\alpha_j - \alpha_{j-1}) 
        \sum_{k = 1}^L  \sum_{i \in \C{A}_j^{(k)}} 
     (\tilde{s}_i^{(k)} - 
   f_{x_i^{(k)}}^{\C{A}_j^c}(\tilde{\ve{s}}_{\C{A}_j^c}(x_i^{(k)})))
   \ge \inf_{\ve{R}: (\ve{R},d) \in \C{RD}_{\rm{in}}(d)}
  \sum_{i=1}^{2^{L-1}} \alpha_i R_i.
$$
Specifically, choose the auxiliary random variables $\ve{u}$
in the Berger-Tung inner bound such that
$$
I(x_i^{(L)};u_i|x^{(L-1)}_{\lfloor \frac{i+1}{2} \rfloor}) = \tilde{s}_i^{(L)}
$$
for each observation $i$.  We will first show by induction that
\begin{equation}
\label{eq:showfirst}
I(x_i^{(k)};\ve{u}|x_{\lfloor (i+1)/2 \rfloor}^{(k-1)})
  = \tilde{s}_i^{(k)}
\end{equation}
for all variables $x_i^{(k)}$ in the tree.
This is true of the leaf variables $x_i^{(L)}$, $i = 1,\ldots,2^{L-1}$
by hypothesis. Next consider a variable $x_i^{(k)}$ and suppose 
the condition holds for $x_{2i-1}^{(k+1)}$ and $x_{2i}^{(k+1)}$.
By the observation in Appendix~\ref{app:equality},
\begin{align*}
I(x_i^{(k)}; \ve{u}|x_{\lfloor (i+1)/2 \rfloor}^{(k-1)})
  & = f_{x_i^{(k)}}(I(x_{2i-1}^{(k+1)};\ve{u}|x_i^{(k)}),
    I(x_{2i}^{(k+1)};\ve{u}|x_i^{(k)})) \\
  & = f_{x_i^{(k)}}(\tilde{s}_{2i-1}^{(k+1)},\tilde{s}_{2i}^{(k+1)}) \\
  & = \tilde{s}_i^{(k)}.
\end{align*}
This establishes~(\ref{eq:showfirst}). Then
$$
\E[(x_1^{(1)} - \E[x_1^{(1)}|\ve{u}])^2] = \sigma^2_{x_1^{(1)}} 
   \exp(-2 \tilde{s}_1^{(1)}) = d.
$$
Thus $\ve{u}$ is in $\C{U}(d)$. If we let
$$
\tilde{R}_i = I(x_i^{(L)}; u_i|u_1,\ldots,u_{i-1}),
$$
then $(\tilde{\ve{R}},d)$ is in $\C{RD}_\mathrm{in}$.
Since $u_i$ is conditionally independent of $\ve{u}$ and all of the
source variables given $x_i^{(L)}$, it follows that
$\tilde{s}_i^{(L)} = 0$ if and only if $u_i$ is independent
of all of the other variables.  We will show that
$$
\sum_{i = j}^{2^{L-1}} \tilde{R}_i =
     I(x_{j}^{(L)},\ldots,x_{2^{L-1}}^{(L)};
        u_j,\ldots,u_{2^{L-1}}|u_1,\ldots,u_{j-1})
$$
by induction. For $j = 2^{L-1}$, this condition holds by the 
definition of $\tilde{R}_j$. Next suppose that the condition
holds for $j$. Then by the tree structure,
\begin{align*}
\sum_{i = j-1}^{2^{L-1}} \tilde{R}_i
  & = I(x_{j-1}^{(L)};u_{j-1}|u_1,\ldots,u_{j-2})
    + I(x_j^{(L)},\ldots,x_{2^{L-1}}^{(L)};
        u_j,\ldots,u_{2^{L-1}}|u_1,\ldots,u_{j-1}) \\
  & = I(x_{j-1}^{(L)},\ldots,x_{2^{L-1}}^{(L)};
              u_{j-1}|u_1,\ldots,u_{j-2})
    + I(x_{j-1}^{(L)},\ldots,x_{2^{L-1}}^{(L)};
        u_j,\ldots,u_{2^{L-1}}|u_1,\ldots,u_{j-1}) \\
  & = I(x_{j-1}^{(L)},\ldots,x_{2^{L-1}}^{(L)};
        u_{j-1},\ldots,u_{2^{L-1}}|u_1,\ldots,u_{j-2}).
\end{align*}
Thus
\begin{align*}
   \inf_{\ve{R}: (\ve{R},d) \in \C{RD}_{\rm{in}}(d)} 
     \sum_{i=1}^{2^{L-1}} \alpha_i R_i
   & \le \sum_{i = 1}^{2^{L-1}} \alpha_i \tilde{R}_i \\
  & \le \sum_{j = 1}^{2^{L-1}} (\alpha_j - \alpha_{j-1}) 
     I(x_j^{(L)},\ldots,x_{2^{L-1}}^{(L)};
    u_j,\ldots,u_{2^{L-1}}|u_1,\ldots,u_{j-1}) \\
  & = \sum_{j = 1}^{2^{L-1}} (\alpha_j - \alpha_{j-1}) 
     I(\ve{x}_{\C{A}_j}^{(L)};
    \ve{u}_{\C{A}_j}|\ve{u}_{\C{A}_j^c}).
\end{align*}
By mimicking (\ref{eq:eqa}) through (\ref{eq:chnrule2}), one can show that
$$
I(\ve{x}^{(L)}_{\C{A}_j}; \ve{u}_{\C{A}_j}|\ve{u}_{\C{A}_j^c})
  = \sum_{k = 1}^L \sum_{i \in \C{A}_j^{(k)}}
  (\tilde{s}_i^{(k)} - I(x_i^{(k)}; \ve{u}_{\C{A}_j^c}|
  x_{\lfloor (i+1)/2 \rfloor}^{(k-1)})).
$$
But by Lemma~\ref{lemma:3.2again} 
and the observation in Appendix~\ref{app:equality},
$$
  I(x_i^{(k)}; \ve{u}_{\C{A}_j^c}|
  x_{\lfloor (i+1)/2 \rfloor}^{(k-1)})
  = f_{x_i^{(k)}}^{\C{A}_j^c}(\tilde{\ve{s}}_{\C{A}_j^c}(x_i^{(k)})).
$$
It follows that
$$
\inf_{\ve{R}: (\ve{R},d) \in \C{RD}_\mathrm{in}}
  \sum_{i = 1}^{2^{L-1}} \alpha_i R_i \le
   \inf_{\ve{R}: (\ve{R},d) \in \C{RD}_\mathrm{out}}
  \sum_{i = 1}^{2^{L-1}} \alpha_i R_i + 2\epsilon.
$$
Since $\epsilon$ was arbitrary, the proof is complete.

\section{Proof of Theorem~\ref{thm:main}}
\label{appendix:cont}

We must show that $\C{RD}^* \subseteq \Co(\C{RD}_\mathrm{in})$.
Since both sets are convex, it suffices to show that for any
componentwise nonnegative vector $(\beta_1,\ldots,\beta_{2^{L-1}},\beta)$
\begin{align}
\label{cont:goal}
\inf_{(\ve{R},d) \in \C{RD}^*} \sum_{i = 1}^{2^{L-1}} \beta_i R_i +
    \beta d
    & \ge
  \inf_{(\ve{R},d) \in \Co(\C{RD}_\mathrm{in})}
     \sum_{i = 1}^{2^{L-1}} \beta_i R_i + \beta d \\
\nonumber
  & =
  \inf_{(\ve{R},d) \in \C{RD}_\mathrm{in}}
     \sum_{i = 1}^{2^{L-1}} \beta_i R_i + \beta d.
\end{align}
We shall assume that $\beta_1 \le \beta_2 \le \cdots \le \beta_{2^{L-1}}$;
the other cases are similar.
Let us temporarily use $\C{RD}^*(\m{K}_x)$ to denote the rate-distortion
region for the binary tree structure problem
when the source variables have covariance matrix $\m{K}_x$ and
similarly for $\C{RD}_\mathrm{in}(\m{K}_x)$. If $\m{K}_x$ is such that
all of the noise variances are positive,
then~(\ref{cont:goal}) follows from Lemma~\ref{lemma:outer}.

If some of the noise variances are zero,
then let $\m{K}^{(n)}_x$ be a sequence of source
covariance matrices converging to $\m{K}_x$ such that for 
each $n$, $\m{K}^{(n)}_x$
corresponds to a source satisfying the binary tree structure for
which all of the noise variances are positive.
Then $\C{RD}^*(\m{K}^{(n)}_x) = \Co(\C{RD}_\mathrm{in}(\m{K}^{(n)}_x))$
for each $n$, so
$$
\inf_{(\ve{R},d) \in \C{RD}^*(\m{K}^{(n)}_x)}
  \sum_{i = 1}^{2^{L-1}} \beta_i R_i +
    \beta d
    =
  \inf_{(\ve{R},d) \in \C{RD}_\mathrm{in}(\m{K}^{(n)}_x)}
     \sum_{i = 1}^{2^{L-1}} \beta_i R_i + \beta d.
$$
We will first show that
\begin{equation}
\label{cont:liminf}
 \liminf_{n \rightarrow \infty}
    \inf_{(\ve{R},d) \in \C{RD}_\mathrm{in}(\m{K}^{(n)}_x)}
     \sum_{i = 1}^{2^{L-1}} \beta_i R_i + \beta d
   \ge
    \inf_{(\ve{R},d) \in \C{RD}_\mathrm{in}(\m{K}_x)}
     \sum_{i = 1}^{2^{L-1}} \beta_i R_i + \beta d.
\end{equation}
For each $n$, there exists a set of auxiliary random
variables $\ve{u}^{(n)}$ such that~\cite[Lemma~3.3]{TH97}
\begin{multline}
\label{vertex:eq}
 \inf_{(\ve{R},d) \in \C{RD}_\mathrm{in}(\m{K}^{(n)}_x)}
     \sum_{i = 1}^{2^{L-1}} \beta_i R_i + \beta d \\
   = \sum_{i = 1}^{2^{L-1}} \beta_i I(u^{(n)}_i;x^{(L,n)}_i|x^{(L,n)}_1,
    \ldots,x_{i-1}^{(L,n)}) +
    \beta \E\lbr\lp x_1^{(1,n)} - \E[x_1^{(1,n)}|
     \ve{u}^{(n)}]\rp^2 \rbr.
\end{multline}
Here $x^{(L,n)}_i$ denotes the $i$th variable at depth $L$ of the
tree corresponding to covariance matrix $\m{K}_x^{(n)}$. Now the
auxiliary random variables $\ve{u}^{(n)}$ can be parametrized by
a compact set, so
consider a subsequence of $\m{K}_x^{(n)}$ along which
$\ve{u}^{(n)}$ converges in distribution to a limit
$\ve{u}$ and the right-hand side of~(\ref{vertex:eq}) converges
to the $\liminf$. Then
\begin{align*}
 & \liminf_{n \rightarrow \infty}
    \inf_{(\ve{R},d) \in \C{RD}_\mathrm{in}(\m{K}^{(n)}_x)}
     \sum_{i = 1}^{2^{L-1}} \beta_i R_i + \beta d
 \\
   & =
   \sum_{i = 1}^{2^{L-1}} \beta_i I(u_i;x^{(L)}_i|x^{(L)}_1,
    \ldots,x_{i-1}^{(L)}) + \beta \E\lbr\lp(x_1^{(1)} - \E[x_1^{(1)}|
     \ve{u}]\rp^2\rbr \\
   & \ge \inf_{(\ve{R},d) \in \C{RD}_\mathrm{in}(\m{K}_x)}
     \sum_{i = 1}^{2^{L-1}} \beta_i R_i + \beta d.
\end{align*}
This establishes~(\ref{cont:liminf}).
On the other hand, Chen and Wagner~\cite{CW07}
have shown that the rate-distortion region is inner-semicontinuous:
$$
  \limsup_{n \rightarrow \infty}
   \inf_{(\ve{R},d) \in \C{RD}^*(\m{K}^{(n)}_x)}
  \sum_{i = 1}^{2^{L-1}} \beta_i R_i +
    \beta d \le
  \inf_{(\ve{R},d) \in \C{RD}^*(\m{K}_x)}
  \sum_{i = 1}^{2^{L-1}} \beta_i R_i +
    \beta d.
$$
Together with~(\ref{cont:liminf}),
this establishes~(\ref{cont:goal}) and hence Theorem~\ref{thm:main}.

\section{Proof of Theorem~\ref{theorem:robust}}
\label{app:robust}

It suffices to show~(\ref{eq:robust2}). If $(\ve{R},d)$ is in
$\C{RD}_\mathrm{in}$, then there exist auxiliary random
variables $\ve{u}$ in $\C{U}(d)$ such that
\begin{align*}
d & \ge \E\Lbr \lp x_1^{(1)} - \E[x_1^{(1)}|\ve{u}]\rp^2 \Rbr \\
\intertext{and}
\sum_{i \in \C{A}} R_i & \ge I\lp\ve{x}_\C{A}^{(L)};\ve{u}_\C{A}|\ve{u}_{\C{A}^c}\rp
\end{align*}
for all $\C{A}$. Now for each $i$, 
$$
u_i = \alpha_i x_i^{(L)} + w_i,
$$
where $w_i$ is Gaussian and independent of $x_i^{(L)}$. Let
$\tilde{u}_i$ be a quantized version of $\tilde{x}_i^{(L)}$
using the same test channel
$$
\tilde{u}_i = \alpha_i \tilde{x}_i^{(L)} + w_i.
$$

Let $\mathrm{MMSE}\lp x_1^{(1)}|\ve{u}\rp$ denote the
mean-square error of the minimum mean-square error
(MMSE) estimate of $x_1^{(1)}$ given $\ve{u}$. Likewise,
let $\mathrm{LLSE}\lp x_1^{(1)}|\ve{u}\rp$ denote the
mean-square error of the linear least-square error
(LLSE) estimate of $x_1^{(1)}$ given $\ve{u}$. Then
\begin{align*}
\E\Lbr \lp \tilde{x}_1^{(1)} - 
  \E[\tilde{x}_1^{(1)}|\tilde{\ve{u}}]\rp^2 \Rbr
  & = \mathrm{MMSE}(\tilde{x}_1^{(1)}|\tilde{\ve{u}}) \\
  & \le \mathrm{LLSE}(\tilde{x}_1^{(1)}|\tilde{\ve{u}}) \\
  & = \mathrm{LLSE}(x_1^{(1)}|\ve{u}) \\
  & = \mathrm{MMSE}(x_1^{(1)}|\ve{u}) \\
  & \le d.
\end{align*}
Also, for any $\C{A}$,
\begin{align*}
\sum_{i \in A} R_i & \ge
   I(\ve{x}_\C{A}^{(L)};\ve{u}_\C{A}|\ve{u}_{\C{A}^c}) \\
   & = h(\ve{u}_\C{A}|\ve{u}_{\C{A}^c}) -
     h(\ve{u}_\C{A}|\ve{u}_{\C{A}^c},\ve{x}_\C{A}^{(L)}) \\
   & = h(\ve{u}_\C{A}|\ve{u}_{\C{A}^c}) -
      h(\ve{u}_\C{A}|\ve{x}_\C{A}^{(L)}) \\
   & \ge h(\tilde{\ve{u}}_\C{A}|\tilde{\ve{u}}_{\C{A}^c}) -
      h(\ve{u}_\C{A}|\ve{x}_\C{A}^{(L)})  \\
   & = h(\tilde{\ve{u}}_\C{A}|\tilde{\ve{u}}_{\C{A}^c}) -
      h(\tilde{\ve{u}}_\C{A}|\tilde{\ve{x}}_\C{A}^{(L)})  \\
   & = h(\tilde{\ve{u}}_\C{A}|\tilde{\ve{u}}_{\C{A}^c}) -
      h(\tilde{\ve{u}}_\C{A}|\tilde{\ve{u}}_{\C{A}^c},\tilde{\ve{x}}_\C{A}^{(L)})  \\
   & = I(\tilde{\ve{x}}_\C{A}^{(L)};
   \tilde{\ve{u}}_\C{A}|\tilde{\ve{u}}_{\C{A}^c})
\end{align*}
where in the inequality we have used the fact
that the Gaussian distribution maximizes entropy for
a fixed covariance.  It follows that $(\ve{R},d)$ is in
$\widetilde{\C{RD}}_\mathrm{in}$.

\section{Proof of Proposition~\ref{embed}}
\label{embedproof}

\begin{figure}[ht]
\begin{center}
\scalebox{1.5}{\input{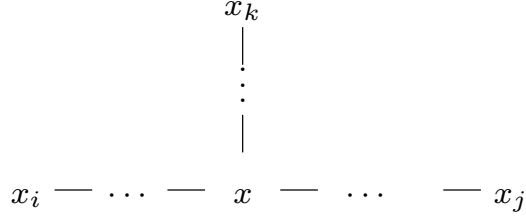}}
\end{center}
\caption{$x$ is the point at which the two paths split.}
\label{paths}
\end{figure}

Suppose that $x_1,\ldots,x_N$ can be embedded in a Gauss-Markov
tree and fix distinct indices $i$, $j$, and $k$.
Without loss of generality, we
may assume that all variables in the tree have mean zero and
variance one.
Consider two paths (i.e., two sequences
of variables), one from $x_i$ to $x_j$ and one from $x_i$ to $x_k$.
Evidently both paths contain $x_i$; let $x$ denote the last variable
in the first path that is contained in the second. This is the
point at which the two paths split, as shown in Fig.~\ref{paths}.
Note that it is possible for $x$ to equal $x_i$, $x_j$, or $x_k$.

Now since $x$ is along the path from $x_i$ to $x_j$, it follows
from the tree condition
that $x_i \markov x \markov x_j$.
Likewise $x_i \markov x \markov x_k$.
Since all of the variables are standard Normals, this
implies~\cite[(5.13)]{WH85}
\begin{align}
\label{gaussmarkov1}
\rho_{ij} & = \E[x_i x] \E[x x_j] \\
\label{gaussmarkov2}
\rho_{ik} & = \E[x_i x] \E[x x_k].
\end{align}

Next consider the paths from $x_j$ to $x_i$ and from
$x_j$ to $x_k$, and let $\tilde{x}$ denote the last
variable in the first path that is contained in the second.
Then both $x$ and $\tilde{x}$ lie along the path from
$x_j$ to $x_i$. If $x \ne \tilde{x}$, then the path
from $x$ to $x_k$ to $\tilde{x}$ to $x$ would form
a loop, which is impossible since the graph is a tree.
Thus $\tilde{x}$ must equal $x$. Thus
$x_j \markov x \markov x_k$ and
$$
\rho_{jk} = \E[x_j x] \E[x x_k].
$$
Combining this equation with~(\ref{gaussmarkov1})
and~(\ref{gaussmarkov2}) yields conditions~(\ref{firstcond})
and~(\ref{secondcond}).

Now suppose that $N = 3$ and conditions~(\ref{firstcond})
and~(\ref{secondcond}) hold. If $\rho_{ij}$ is nonzero
for all $i \ne j$, then
$$
0 < \frac{\rho_{ij} \rho_{ik}}{\rho_{jk}} \le 1
$$
for all distinct $i$, $j$, and $k$. This implies
that $x_1$, $x_2$, and $x_3$, can be written
\begin{align*}
x_1 & = \sqrt{\frac{\rho_{12}\rho_{13}}{\rho_{23}}} \cdot
   \sgn(\rho_{23}) \cdot x_0 + z_1 \\
x_2 & = \sqrt{\frac{\rho_{12}\rho_{23}}{\rho_{13}}} \cdot
   \sgn(\rho_{13}) \cdot x_0 + z_2 \\
x_3 & = \sqrt{\frac{\rho_{13}\rho_{23}}{\rho_{12}}} \cdot
   \sgn(\rho_{12}) \cdot x_0 + z_3,
\end{align*}
where $\sgn(\cdot)$ is the signum function
$$
\sgn(\rho) = \begin{cases}
1 & \text{if $\rho > 0$} \\
0 & \text{if $\rho = 0$} \\
-1 & \text{if $\rho < 0$},
\end{cases}
$$
and where $x_0$, $z_1$, $z_2$, $z_3$ are independent Gaussian
random variables. Here $x$ is a standard Normal and
the variances of the $z$s are chosen to such that the
$x$s have unit variance. It is readily verified that
this construction yields the correct correlation coefficients
among the $x$s. It is then clear that $x$ and the $x$s can
be arranged in the Gauss-Markov tree shown in Figure~\ref{xtree}.

If, say, $\rho_{12} = 0$, then by condition~(\ref{firstcond}),
either  $\rho_{13} = 0$ or $\rho_{23} = 0$. Suppose that
$\rho_{13} = 0$. Then $x_1$ is uncorrelated, and hence
independent, of $x_2$ and $x_3$. It follows that the $x$s
can be written
\begin{align*}
x_1 & = z_1 \\
x_2 & = \sqrt{|\rho_{23}|} \cdot x_0 + z_2 \\
x_3 & = \sqrt{|\rho_{23}|} \cdot \sgn(\rho_{23}) \cdot x_0 + z_3,
\end{align*}
so that the  $x_0$ and the $x$s can again
be arranged in the Gauss-Markov tree shown in
Figure~\ref{xtree}.

\section{Proof of Proposition~\ref{BTisbad}}
\label{BTbadproof}

Since we are assuming that $R_3 = 0$, the problem
effectively reduces to a two-encoder setup. By Lemma~\ref{lemma:BT}
and~(\ref{eq:contrapoly3}), the minimum $R_1 + R_2$ equals
\begin{align*}
\text{inf} & \quad I(\ve{x};\ve{u}) \\
\text{subject to} & \quad u_1 \markov x_1 \markov x_2 \markov u_2 \\
             & \quad (\ve{x}, \ve{u}) \ \text{jointly Gaussian} \\
  &  \quad  \E[(x_3 - \E[x_3|\ve{u}])^2] \le d.
\end{align*}
Without loss of generality, we may assume that
\begin{align*}
u_1 & = x_1 + z_1 \\
u_2 & = x_2 + z_2
\end{align*}
where the $z$ variables are Gaussian and independent of each
other and $\ve{x}$. Let $z_1$ have variance $\alpha \sigma^2$
and $z_2$ have variance $\beta \sigma^2$.

Via straightforward calculations one can show that
\be
\label{BTMI}
I(\ve{x};\ve{u}) = \frac{1}{2} \log \left( (1-\rho^2) \alpha^{-1} \beta^{-1}
   + \alpha^{-1} + \beta^{-1} + 1 \right)
\ee
and
$$
\E[(x_3 - \E[x_3|\ve{u}])^2] = 1 - \frac{1}{\sigma^2}
  \frac{2(1 + \rho) + \alpha + \beta}{
    4 (1 + \alpha)(1 + \beta) - 4\rho^2}.
$$
Now
\begin{align*}
\frac{2(1 + \rho) + \alpha + \beta}{
    4(1 + \alpha)(1 + \beta) - 4\rho^2}
   & \le \frac{4 + \alpha + \beta}
     {4\alpha + 4\beta + 4 \alpha \beta} \\
  & \le \frac{1 + \alpha + \beta}{\alpha 
   + \beta + \alpha \beta} \\
  & \le \frac{1}{\alpha \beta} + 2.
\end{align*}
It follows that as $\sigma^2$ tends to infinity, in order to
continue to meet the distortion constraint, we require that
$\alpha \beta$ tend to zero. But this implies that
$I(\ve{x};\ve{u})$ tend to infinity, by~(\ref{BTMI}).

\section{Proof of Proposition~\ref{latticegood}}
\label{latticeproof}

Since the average distortion
is the same for all $\ell$, let us assume that $\ell = 1$ and
write $x_3$ in place of $x_3(1)$ and likewise for the other
variables. Then by the triangle inequality
\begin{align*}
\sqrt{\E[(x_3 - \hat{x}_3)^2]} 
   \le \sqrt{\E[(x_3 - (\tilde{x}_1 - \tilde{x}_2))^2]}
      + \sqrt{\E[((\tilde{x}_1 - \tilde{x}_2) - \hat{x}_3)^2]}.
\end{align*}
Now
$$
|x_1 - \tilde{x}_1| \le 2^{-(n + 1)}
$$
and likewise for $|x_2 - \tilde{x}_2|$. Thus
$$
\E[(x_3 - (\tilde{x}_1 - \tilde{x}_2))^2] \le 2^{-2n}.
$$
Define the event
$$
A = \{|\tilde{x}_1 - \tilde{x}_2| < 2^{m-1} \}.
$$
Now on $A$, 
\begin{align*}
\hat{x}_3 & = u_1 - u_2 \mod \Lambda_o \\
       & = \tilde{x}_1 - \tilde{x}_2 \mod \Lambda_o \\
     & = \tilde{x}_1 - \tilde{x}_2,
\end{align*}
so
\begin{align*}
\E[(\tilde{x}_1 - \tilde{x}_2 - \hat{x}_3)^2]
  & = \E[(\tilde{x}_1 - \tilde{x}_2 - \hat{x}_3)^2 1_{A^c}] \\
  & \le \sqrt{\E[(\tilde{x}_1 - \tilde{x}_2 - \hat{x}_3)^4] \prob(A^c)}.
\end{align*}
But
\begin{align*}
|\tilde{x}_1 - \tilde{x}_2 - \hat{x}_3| & \le 
   |x_1 - x_2| + |\tilde{x}_1 - x_1| + |x_2 - \tilde{x}_2|
   + |\hat{x}_3| \\
   & \le |x_1 - x_2| + 2^{-n} + 2^{m-1} \\
   & \le |x_1 - x_2| + 2^{m}.
\end{align*}
Since $x_1 - x_2$ is a standard Normal random variable, 
$\E[(x_1 - x_2)^4] = 3$, and Minkowski's inequality implies
$$
\E[(\tilde{x}_1 - \tilde{x}_2 - \hat{x}_3)^4] \le 3 + 2^m.
$$
It only remains to bound $\prob(A^c)$. Using a well-known upper bound
on the tail of the Gaussian distribution
$$
\prob(A^c) \le 2 \exp(-2^{2m-3}).
$$
Combining these various bounds gives
$$
\E[(x_3 - \hat{x}_3)^2] \le (2^{-n} + (2(3 + 2^m) \exp(-2^{2m-3}))^{1/2})^2
$$
Proposition~\ref{latticegood} follows.

\section{Proof of Proposition~\ref{prop:converse}}
\label{app:converse}

Recall we may assume that all of the variables have unit variance.
By Proposition~\ref{embed}, if $x_1$, $x_2$, and $x_3$ cannot be 
embedded in a Gauss-Markov tree, then either
\beq
\label{eq:fail1}
\rho_{12} \rho_{13} \rho_{23} < 0
\eeq
or
\beq
\label{eq:fail2}
|\rho_{ij}| < |\rho_{ik} \rho_{kj}|
\eeq
for some distinct $i$, $j$, and $k$.

Suppose first that (\ref{eq:fail1}) 
holds. 
Then we must have
$|\rho_{ij}| < 1$ for all $i \ne j$.
Now
\begin{align*}
\E[x_1|x_2,x_3] & = \frac{\rho_{12} - \rho_{13} \rho_{23}}{1 - \rho_{23}^2} 
    \cdot x_2
    + \frac{\rho_{13} - \rho_{12} \rho_{23}}{1 - \rho_{23}^2} \cdot x_3 \\
   & \df a_2 x_2 + a_3 x_3.
\end{align*}
Then
\begin{equation}
\label{eq:prodform}
a_2 \cdot a_3 \cdot \rho_{23} =
  \frac{1}{(1 - \rho_{23}^2)^2} \frac{\rho_{23}}{\rho_{13} \rho_{12}}
  (\rho_{12}^2 - \rho_{12} \rho_{13} \rho_{23}) (\rho_{13}^2 - 
    \rho_{12} \rho_{13} \rho_{23}),
\end{equation}
which is negative by~(\ref{eq:fail1}).
This establishes the desired conclusion in this case. We will
therefore assume throughout the remainder of the proof
that $\rho_{12} \rho_{13} \rho_{23} \ge 0$.

Suppose that (\ref{eq:fail2}) holds, say, for $i = 1$, $j = 2$,
and $k = 3$. Then we must have $|\rho_{12}| < 1$ and
$\rho_{13} \cdot \rho_{23} \ne 0$. Furthermore,
if $|\rho_{23}| = 1$, then $|\rho_{12}| = |\rho_{13}|$, which
would contradict~(\ref{eq:fail2}).
Thus we may assume that $|\rho_{23}| < 1$.
First suppose that $\rho_{12} = 0$.  Then
$$
a_2 \cdot a_3 \cdot \rho_{23} = - \frac{\rho^2_{13} \rho^2_{23}}
    {(1 - \rho_{23}^2)^2},
$$
which is negative.
We will therefore focus on the
case in which $\rho_{12} \rho_{13} \rho_{23} > 0$.

Next observe that since we are assuming that (\ref{eq:fail2}) holds
for $i = 1$, $j = 2$, and $k = 3$, the opposite inequality
must hold strictly in the other two cases
\begin{align*}
|\rho_{13}| & > |\rho_{12} \rho_{23}| \\
|\rho_{23}| & > |\rho_{12} \rho_{13}|.
\end{align*}
This can be seen by contradiction: if, e.g., $|\rho_{13}| \le
|\rho_{12} \rho_{23}|$, then combining this fact with
(\ref{eq:fail2}) yields
$$
|\rho_{12}| < |\rho_{13} \rho_{23}| \le |\rho_{12}| |\rho_{23}|^2
$$
which is evidently false. From (\ref{eq:prodform}) and 
the three assumed conditions, $\rho_{12} \rho_{13} \rho_{23} > 0$,
$|\rho_{12}| < |\rho_{13} \rho_{23}|$, and
$|\rho_{13}| > |\rho_{12} \rho_{23}|$, it follows that
$a_2 \cdot a_3 \cdot \rho_{23}$ is
negative, as desired.


\begin{thebibliography}{99}

\bibitem{B78} T.~Berger, {\em Multiterminal Source Coding}, series: The
Information Theory Approach to Communications, Vol.~229, CISM courses
and lectures, Springer-Verlag, 1978.

\bibitem{CW07} J.~Chen and A.~B.~Wagner, ``Inner Semicontinuity of
  Gaussian Rate-Distortion Regions with Applications,'' preprint.

\bibitem{CZBW04} J.~Chen, X.~Zhang, T.~Berger, S.~B.~Wicker,
   ``An upper bound on the sum-rate distortion function
                   and its corresponding rate allocation schemes for
                     the {CEO} problem,'' \emph{IEEE Transactions on 
       Information Theory}, v.~22, No.~6, Aug., 2004, pp.~977--987.

\bibitem{HT97} S. Hanly and D. Tse, ``Multi-Access Fading Channels:
  Part II: Delay-Limited Capacities'', {\em IEEE Transactions on Information
  Theory}, v. 44, No. 7, Nov., 1998, pp. 2816-2831.

\bibitem{KP07} D.~Krithivasan and S.~S.~Pradhan, ``Lattices for Distributed
   Source Coding: Jointly Gaussian Sources and Reconstruction of a Linear
   Function,'' \texttt{arXiv:0707.3461}.

\bibitem{Oohama97} Y.~Oohama, ``{G}aussian Multiterminal Source Coding,''
{\em IEEE Transactions on Information Theory},
Vol.~43(6), pp.~1912-1923, Nov., 1997.

\bibitem{Oohama05} Y.~Oohama, ``Rate-Distortion Theory for Gaussian Multiterminal
Source Coding Systems with Several Side Informations at the
Decoder", {\em IEEE Transactions on Information  Theory},
Vol.~51(7), pp.~2577-2593, July 2005.

\bibitem{Oohama06} Y.~Oohama, ``Gaussian Multiterminal Source Coding with Several
Side Informations at the Decoder", {\em IEEE Symposium on
Information Theory}, 2006.

\bibitem{PTR04} V.~Prabhakaran, D.~Tse and K.~Ramchandran, ``Rate Region
of the Quadratic Gaussian CEO Problem", {\em IEEE Symposium  on Information
Theory}, 2004.

\bibitem{SK86} T.~P.~Speed and H.~T.~Kiiveri, ``Gaussian Markov
   Distributions Over Finite Graphs,'' {\em Annals of Statistics},
    Vol.~14(1), pp.~138-150, Mar., 1986.

\bibitem{TH97} D. Tse and S. Hanly, ``Multi-Access Fading Channels:
  Part I: Polymatroid Structure, Optimal Resource Allocation and
  Throughput Capacities'', {\em IEEE Transactions on Information Theory},
  Vol.~44(7), Nov.~1998, pp. 2796-2815.

\bibitem{Tung:PHD} S.-Y.~Tung, ``Multiterminal Source Coding", Ph.D.\ dissertation,
Cornell University, 1978.

\bibitem{WTV06} A.\ B.\ Wagner, S.\ Tavildar and P.~Viswanath,  ``Rate Region of
the Quadratic Gaussian Two-Terminal Source-Coding Problem",
submitted to {\em IEEE Transactions on Information Theory}, accepted.

\bibitem{WV06} H. Wang and P. Viswanath,
``Vector Gaussian Multiple Description for Individual and Central
Receivers", {\em IEEE Transactions on Information Theory},
Vol.~53(6), pp.~2133-2153, June 2007.

\bibitem{Matroid} D. J. A. Welsh,  {\em Matroid theory}, Academic Press,
  London, 1976.

\bibitem{WH85} E.~Wong and B.~Hajek {\em Stochastic Processes in
   Engineering Systems}, Springer-Verlag, New York, 1985.

\end{thebibliography}
\end{document}